\documentclass[11pt]{article}

\usepackage[T1]{fontenc}

\usepackage{lineno,hyperref}
\modulolinenumbers[5]

\usepackage{amsmath}
\usepackage{amssymb}
\usepackage{amsthm}
\usepackage{tikz}
\usepackage{pgfplots}
\usepackage{graphicx}
\usepackage{slashbox}
\usepackage{caption}
\usepackage{subcaption}
\usepackage{xcolor}
\usepackage{balance}
\allowdisplaybreaks
\usepackage{algorithm}
\usepackage[noend]{algcompatible}
\usepackage{mathtools}
\usepackage{subcaption}
\usepackage{hyperref}
\usepackage{comment}
\usepackage{tabularx}
\usepackage{bbm}
\usepackage{thmtools}
\usepackage{cite}

\usepackage{hyperref}
\usepackage{pgfplots}
\usepackage{graphicx}
\usepackage{caption}
\usepackage{subcaption}
\usepackage{algorithm}
\usepackage[noend]{algcompatible}
\usepackage{multirow}
\usepackage{booktabs}
\usepackage{enumitem}
\usepackage{array}
\usepackage{authblk}
\usepackage{setspace}
\usepackage{fullpage,etoolbox}
\usepackage{color, colortbl}
\definecolor{Gray}{gray}{0.9}
\usepackage[most,breakable,many]{tcolorbox}
\tcbset{before skip=0pt,after skip=0pt} 

\allowdisplaybreaks

\renewcommand{\baselinestretch}{0.97}

\begin{document}
\newtheorem{definition}{Definition} 
\newtheorem{theorem}{Theorem} 
\newtheorem{assumption}{Assumption} 
\newtheorem{problem}{Problem} 
\newtheorem{remark}{Remark}
\newtheorem{lemma}{Lemma}
\newtheorem{corollary}{Corollary}
\newtheorem{proposition}{Proposition}
\newtheorem{example}{Example}

 \title{Robust Conformal Prediction for STL Runtime Verification \\under Distribution Shift}

\author[1]{Yiqi Zhao}
\author[2]{Bardh Hoxha}
\author[2]{Georgios Fainekos}
\author[1]{Jyotirmoy V. Deshmukh}
\author[1]{Lars Lindemann}
\affil[1]{Thomas Lord Department of Computer Science, University of Southern California}
\affil[2]{Toyota NA R\&D}

\date{\vspace{-5ex}}

\providecommand{\keywords}[1]{\textbf{\textit{Index terms---}} #1}

\maketitle

\thispagestyle{plain}
\pagestyle{plain}
\renewcommand{\baselinestretch}{0.97}

\begin{abstract}
Cyber-physical systems (CPS) designed in simulators behave differently in the real-world. Once they are deployed in the real-world, we would hence like to predict system failures during runtime. We propose robust predictive runtime verification (RPRV) algorithms under signal temporal logic (STL) tasks for general stochastic CPS. The RPRV problem faces several challenges: (1) there may not be sufficient data of the behavior of the deployed CPS, (2) predictive models are based on a  distribution over system trajectories encountered during the design phase, i.e., there may be a distribution shift during deployment. To address these challenges, we assume to know an upper bound on the statistical distance (in terms of an f-divergence) between the distributions at  deployment and design time, and we utilize techniques based on {\em robust conformal prediction}. 
 Motivated by our results in \cite{lindemann2023conformal}, 
we construct an accurate and an interpretable RPRV algorithm. We use a trajectory prediction model to estimate the system behavior at runtime and robust conformal prediction to obtain probabilistic guarantees by accounting for distribution shifts. We precisely quantify the relationship between calibration data, desired confidence, and permissible distribution shift. To the best of our knowledge, these are the first statistically valid algorithms under distribution shift in this setting. We empirically validate our algorithms on a Franka manipulator within the NVIDIA Isaac sim environment.

\end{abstract}


\section{Introduction}
\label{sec:introduction}

Cyber-physical Systems (CPS) operate in highly unpredictable environments and often have intrinsic and extrinsic sources of stochasticity that affect their dynamic behavior. Many such CPSs can be modeled simply as a distribution $\mathcal{D}$ over the space of system trajectories. In this paper, we are interested in the predictive runtime verification of such a stochastic CPS against a specification $\phi$ expressed in signal temporal logic (STL). In other words, during the operation of the system we would like to compute the probability that the system will satisfy (or violate) the specification $\phi$ using the already observed part of the system trajectory. System verification using formal languages is of great interest for the safe design of many CPS applications, e.g., the responsibility sensitive safety  model can be encoded in STL for verification of safe behaviors in autonomous vehicles  \cite{hekmatnejad2019encoding}. 

The use of high-fidelity models and  simulators has resulted in statistical and data-driven paradigms for the design, analysis, and test of many CPS. While previous approaches have formulated algorithms for predictive runtime monitoring of CPSs \cite{lindemann2023conformal,cairoli2023conformal}, the guarantees that these approaches provide depend on the underlying distribution being the same at test and design time. However, when a CPS  is deployed in the real-world, the underlying distribution may be different -- a phenomenon called a {\em distribution shift}.
Our focus is thus on designing \emph{robust predictive runtime verification (RPRV) algorithms} that provide valid verification results even when the test distribution $\mathcal{D}$ is different from the training distribution~$\mathcal{D}_0$.

\begin{figure}
    \centering
    \includegraphics[scale=0.3]{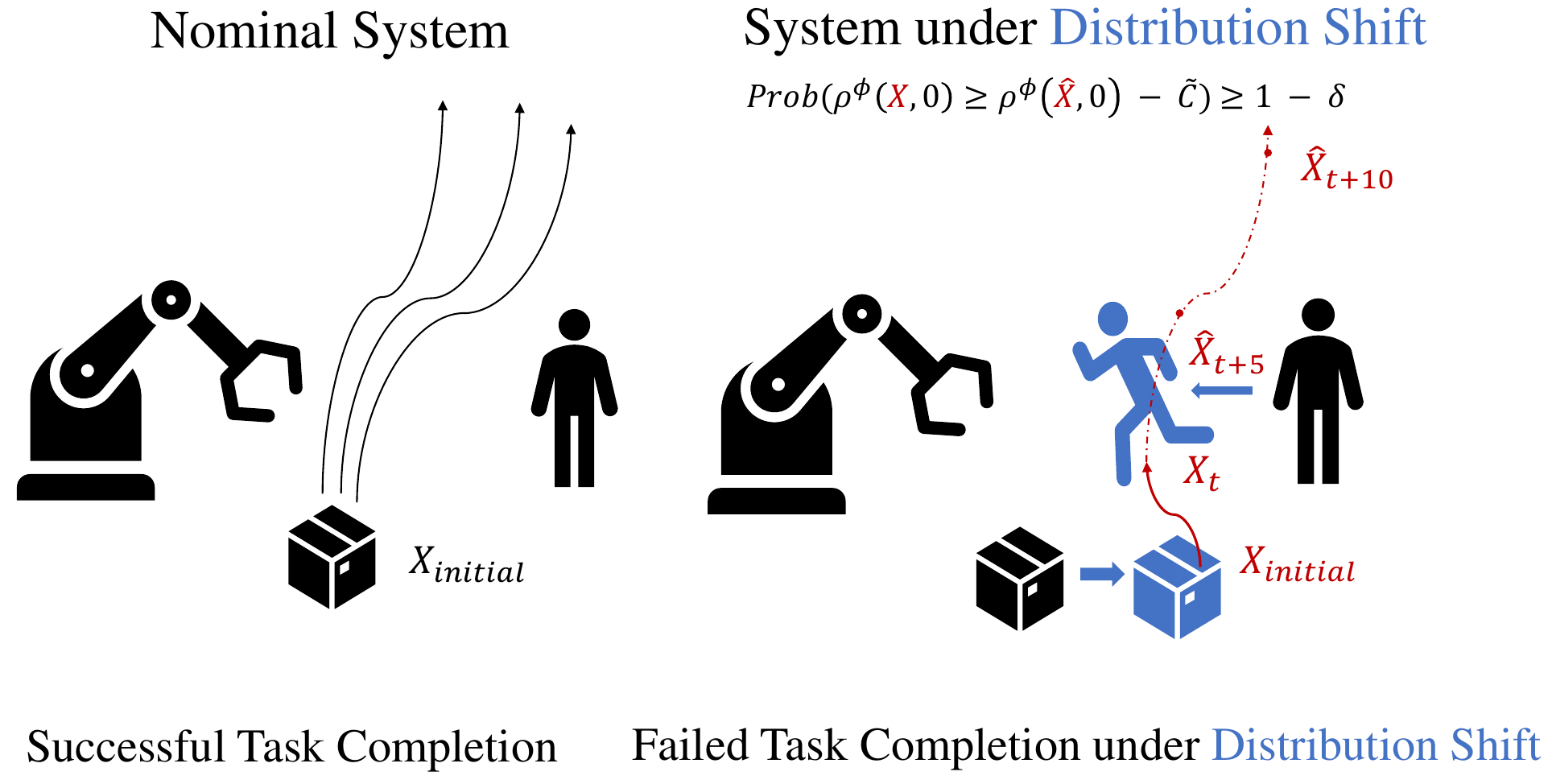}
    \caption{Left: A robotic arm manipulating  a box in the presence of a human. Right: The robotic arm performing the same task when the human behavior and the package location changed. We propose predictive robust runtime verification algorithms to verify systems under such distribution shift.}
    \label{fig:robust_stl}
\end{figure}

Specifically, we propose RPRV algorithms that account for all test distributions $\mathcal{D}$ that are contained within a set of possible test distributions $P(\mathcal{D}_0)$. The set $P(\mathcal{D}_0)$ here denotes all distributions that are close to the training distribution $\mathcal{D}_0$ under a suitable distance measure. For instance, $P(\mathcal{D}_0)$ may denote those distributions $\mathcal{D}$ whose f-divergence with respect to $\mathcal{D}_0$ is bounded by $\epsilon>0$, i.e., $P(\mathcal{D}_0)$ may be defined as $\{\mathcal{D}|D_f(\mathcal{D},\mathcal{D}_0)\le \epsilon\}$ where $D_f$ is an f-divergence measure. RPRV algorithms are needed since test distributions are usually different from the training distribution. Examples include varying conditions in the environment that the system operates in, e.g., weather conditions or traffic, or also when $\mathcal{D}_0$ models a high-fidelity simulator. To the best of our knowledge, existing predictive runtime verification algorithms are not robust to distribution shift. Furthermore, in this paper, we deal with the challenge of increasing data requirements for larger distribution shifts. We make the following contributions.


\begin{itemize}
\item We present an accurate and an interpretable RPRV algorithms that i) use trajectory predictors to predict future system behavior, and ii) leverage robust conformal prediction \cite{cauchois2020robust} to quantify prediction uncertainty using calibration data from  $\mathcal{D}_0$. The algorithms are valid with a user-defined probability for all test distributions $\mathcal{D}\in P(\mathcal{D}_0)$. 
\item We analyze data requirements and the interplay between confidence and permissible distribution shift. We make various algorithmic advancements of the interpretable method, originally presented in \cite{lindemann2023conformal}, to improve accuracy and data efficiency for increasing distribution shifts.  
\item We empirically validate our algorithms on a stylized running example and a Franka manipulator within NVIDIA Isaac sim. We illustrate the efficacy of our robust algorithms compared to a non-robust version from \cite{lindemann2023conformal}. 
\end{itemize}

\subsection{Related Work.} 

\noindent \textbf{(Offline) Verification of Stochastic Systems.} 
Formal verification of general CPS models is known to be an undecidable problem \cite{asarin1998achilles}.
To address the computational cost and inefficiency of abstraction-based verification algorithms, techniques such as statistical model checking have been used to give statistical guarantees \cite{agha2018survey,legay2019statistical,zarei2020statistical}.
In \cite{sen2004statistical}, statistical hypothesis testing is performed with a set of executions from a black-box system to verify continuous stochastic logic specifications. Statistical verification under STL specifications was first considered in \cite{bartocci2013robustness,bartocci2015system}. The works in \cite{zarei2020statistical,qin2022statistical} consider the problem of statistical verification of learning-enabled CPS with respect to STL specifications. Recent work has also sought to combine model-based techniques and statistical, data-driven techniques \cite{salamati2021data,salamati2020data}. For instance, in \cite{pedrielli2023part} surrogate Gaussian process models of the system are learned and used to to obtain probabilistic guarantees on satisfying STL specifications. These works assume that the distribution from which data is sampled is fixed. The authors in \cite{dutta2023distributionally} propose an active sampling approach using imprecise neural networks to promote distributional robustness in statistical verification of neural networks. In another direction, recent frameworks train generative models that capture distribution shifts \cite{wu2023toward,wong2020learning}.
\\[12pt]
\noindent \textbf{Runtime Verification of Stochastic Systems.} In runtime verification (RV), we are instead interested in verifying system properties during the operation of the system  solely based on the currently observed system trajectory \cite{bartocci2018introduction,bauer2011runtime,lukina2021into,jaeger2020statistical}. RV techniques complement offline verification techniques, and there is a growing body of literature on RV approaches for STL specifications \cite{ma2021predictive, deshmukh2017robust, dokhanchi2014line,gressenbuch2021predictive}. Predictive RV is a special class of RV where we use a system model to predict the future system behavior from the currently observed system trajectory to either check if (hidden) system states satisfy a given specification \cite{sistla2008monitoring,sistla2011runtime} or if the system may violate system specifications in the future \cite{bortolussi2019neural,qin2020clairvoyant,cairoli2023learning,yu2024modeljournal, yu2022online, wang2023sleep}. In our prior work \cite{lindemann2023conformal}, we presented two predictive RV algorithms that use conformal prediction, a statistical tool for uncertainty quantification \cite{vovk2005algorithmic},  by calibrating prediction errors of a trajectory predictor to obtain valid probabilistic verification guarantees on the satisfaction of STL specifications. Similar in spirit, the authors in \cite{cairoli2023conformal} used a technique known as conformalized quantile regression \cite{romano2019conformalized} to design predictive RV algorithms that also provide probabilistic verification guarantees on the satisfaction of STL specifications. While statistical guarantees are provided in \cite{cairoli2023conformal, lindemann2023conformal}, these guarantees are not valid when the trajectory distribution of the deployed system deviates from the distribution of the design-time system.  The authors in \cite{mao2023safe} take a first step in this direction with conformal prediction by proposing robust evaluators that minimize distribution shifts in high-dimensional measurements. However, to the best of our knowledge, no existing work provides statistically valid RV guarantees under distribution shift as we do in this paper.

%
%


\section{Problem Formulation}
\label{sec:rnn}

To describe stochastic systems, we consider an unknown test distribution $\mathcal{D}$ over system trajectories $X:=(X_0,X_1\hdots)\sim \mathcal{D}$ where $X_\tau \in \mathbb{R}^n$ is the state of the system at time $\tau$. We make no assumption about $\mathcal{D}$, e.g.,  $\mathcal{D}$ can  describe Markov decision processes or hybrid stochastic systems. While the distribution $\mathcal{D}$ is completely unknown, we assume that we have access to $K$ calibration trajectories $(X^{(1)},\hdots,X^{(K)})$ from a training distribution $\mathcal{D}_0$ that is close to $\mathcal{D}$ (as specified later).\footnote{The distributions $\mathcal{D}$ and $\mathcal{D}_0$ are  defined over the same  probability space $(\Omega,\mathcal{F},\mathbb{P})$ where $\Omega$ is
the sample space, $\mathcal{F}$ is a $\sigma$-algebra of $\Omega$, and $\mathbb{P}:\mathcal{F}\to[0,1]$ is a probability 
measure. For simplicity, we will mostly use the notation $\text{Prob}$ to be independent of the underlying probability space.}
\begin{assumption}\label{ass1}
 	We have access to a dataset $S:=(X^{(1)},\hdots,X^{(K)})$ in which each of the $K$ trajectories $X^{(i)}:=(X_0^{(i)},X_1^{(i)},\hdots)$ is independently drawn from a training distribution $\mathcal{D}_0$, i.e., $X^{(i)}\sim\mathcal{D}_0$.\footnote{For instance, we can obtain such i.i.d. trajectories from a simulator that we can query repeatedly with a fixed distribution over simulation parameters.}
 \end{assumption}
 Assumption \ref{ass1} holds in many  applications. An example would be a setting in which $\mathcal{D}_0$ describes the motion of a robot within a high-fidelity simulator, while $\mathcal{D}$ describes a real robot operating in a lab environment. To measure closeness of the distributions $\mathcal{D}_0$ and $\mathcal{D}$, we use the f-divergence, a statistical distance, that quantifies the similarity between $\mathcal{D}_0$ and $\mathcal{D}$ and thereby the distribution shift. Specifically, the f-divergence $D_f(\mathcal{D},\mathcal{D}_0)$ is defined as
 \begin{align*}
     D_f(\mathcal{D},\mathcal{D}_0):=\int_\mathcal{X} f\Big(\frac{d \mathcal{D}}{d \mathcal{D}_0}\Big) d \mathcal{D}_0
 \end{align*}
  where $\mathcal{X}$ is the support of $\mathcal{D}_0$, $\mathcal{D}$ is absolutely continuous with respect to $\mathcal{D}_0$, and $\frac{d \mathcal{D}}{d \mathcal{D}_0}$ is the Radon-Nikodym derivative of $\mathcal{D}$ with respect to $\mathcal{D}_0$.  The function $f:[0,\infty)\to\mathbb{R}$ is convex and satisfies $f(1)=0$. If we set $f(z) := \frac{1}{2}|z - 1|$, we obtain the total variation distance  $TV(\mathcal{D}, \mathcal{D}_0) := \frac{1}{2}\int_x|P(x) - Q(x)|dx$ where $P$ and $Q$ are probability density functions corresponding to $\mathcal{D}$ and $\mathcal{D}_0$.
 
\begin{assumption}\label{ass2}
 	The test and training distributions $\mathcal{D}$ and $\mathcal{D}_0$ are such that $D_f(\mathcal{D},\mathcal{D}_0)\le \epsilon$ where $\epsilon>0$. We hence assume that $\mathcal{D}\in P(\mathcal{D}_0):=\{\mathcal{D}' \mid D_f(\mathcal{D}',\mathcal{D}_0)\le \epsilon\}$.
 \end{assumption}
 
We emphasize that the parameter $\epsilon$ is a measure of the permissible distribution shift in terms of the f-divergence $D_f$.
\\[12pt]
\noindent\textbf{Challenges in Runtime Verification.} Given a specification $\phi$ and a partial observation $(X_0, \hdots, X_t)$ from the test trajectory $X\sim \mathcal{D}$ at runtime $t$, we are interested in computing the probability that $X$ satisfies $\phi$. The challenges are that we only have knowledge about the training distribution $\mathcal{D}_0$ as per Assumption \ref{ass1}, and our knowledge about the test distribution $\mathcal{D}$ is limited to knowing that $\mathcal{D}_0$ and $\mathcal{D}$ are $\epsilon$-close as per Assumption \ref{ass2}. We design RPRV algorithms that are predictive in the sense that we use predictions $\hat{X}_{\tau|t}$ of future states  $X_\tau$ for $\tau>t$ and  robust as we provide valid probabilistic guarantees as long as  $\mathcal{D}\in P(\mathcal{D}_0)$.



\subsection{Signal Temporal Logic}
\label{sec:STL}
We use signal temporal logic (STL) to express  system specifications and define STL over discrete-time trajectories $x:=(x_0,x_1,\hdots)$, e.g., $x$ can be a realization of the stochastic trajectory $X$. We note that readers with limited background in temporal logics can, if they like to, skip the following formal definitions of syntax and semantics of an STL formula $\phi$ and instead think of $\phi$ as a high-level system specification that is imposed on the system at time $\tau_0$. We let $(x,\tau_0)\models \phi$ indicate that $x$ satisfies $\phi$ and we assume that bounded trajectories $x$ of length $L^\phi$ are sufficient to compute $(x,\tau_0)\models \phi$. The notation $\rho^\phi(x,\tau_0)\in\mathbb{R}$ will indicate how well $\phi$ is satisfied by $x$ at time $\tau_0$ with larger values indicating better satisfaction.

Formally, the atomic elements of STL are predicates that are functions $\pi:\mathbb{R}^n\to\{\text{True},\text{False}\}$. The predicate $\pi$ is defined via a predicate function $h:\mathbb{R}^n\to\mathbb{R}$ as $\pi(x_\tau):=\text{True}$ if $h(x_\tau)\ge 0$ and $\pi(x_\tau):=\text{False}$ otherwise. The syntax of STL is recursively defined as 
\begin{align}\label{eq:full_STL}
\phi \; ::= \; \text{True} \; | \; \pi \; | \;  \neg \phi' \; | \; \phi' \wedge \phi'' \; | \; \phi'  U_I \phi''
\end{align}
where $\phi'$ and $\phi''$ are STL formulas. The Boolean operators $\neg$ and $\wedge$ encode negations (``not'') and conjunctions (``and''), respectively. The until operator $\phi' {U}_I \phi''$ encodes that $\phi'$ has to be true from now on until $\phi''$ becomes true at some future time within the time interval $I\subseteq \mathbb{R}_{\ge 0}$. We can further derive the operators for disjunction ($\phi' \vee \phi'':=\neg(\neg\phi' \wedge \neg\phi'')$), eventually ($F_I\phi:=\top U_I\phi$), and always ($G_I\phi:=\neg F_I\neg \phi$).

To determine if a trajectory $x$ satisfies an STL formula $\phi$ that is enabled at time $\tau_0$, we can define the semantics as a relation $\models$, i.e.,  $(x,\tau_0) \models\phi$ means that $\phi$ is satisfied. While the STL semantics are fairly standard \cite{maler2004monitoring}, we recall them in Appendix \ref{app:STL}. Additionally, we can define robust semantics $\rho^{\phi}(x,\tau_0)\in\mathbb{R}$ that indicate how robustly the formula $\phi$ is satisfied or violated \cite{donze2,fainekos2009robustness}, see Appendix \ref{app:STL}. Larger and positive values of $\rho^{\phi}(x,\tau_0)$ hence indicate that the specification is satisfied more robustly. Importantly, it holds that $(x,\tau_0)\models \phi$ if $\rho^\phi(x,\tau_0)>0$ due to \cite[Proposition 16]{fainekos2009robustness}. We make the following assumption on the class of STL formulas in this paper.
\begin{assumption}\label{ass3}
  We consider bounded STL formulas $\phi$, i.e., all time intervals $I$ within the formula $\phi$ are bounded.
 \end{assumption}
 
 Satisfaction of bounded STL formulas can be decided by finite length trajectories \cite{sadraddini2015robust}. The minimum length is given by the formula length $L^\phi$, i.e., with knowledge of $(x_{\tau_0},\hdots,x_{\tau_0+L^\phi})$ we can compute $(x,\tau_0) \models\phi$, see again Appendix \ref{app:STL} for more details.

\subsection{Robust Predictive Runtime Verification}\label{sec:robust_cp}

Assume that we have observed the states $X_{\text{obs}}:=(X_0,\hdots,X_t)$ at runtime $t$, i.e., all states up until time $t$ are known, while future states $X_{\text{un}}:=(X_{t+1},X_{t+2},\hdots)$ from ${X}=(X_{\text{obs}},X_{\text{un}})\sim \mathcal{D}$ are not known yet. In this paper, we are interested in solving the following problem. 
  
 \begin{problem}\label{prob1}
Let $\mathcal{D}_0$ be a training distribution, $S$ be a calibration dataset from $\mathcal{D}_0$ that satisfies Assumption \ref{ass1}, $\mathcal{D}$ be a test distribution from $P(\mathcal{D}_0)$ that satisfies Assumption \ref{ass2}, and $\phi$ be an STL formula imposed at time $\tau_0$ that satisfies Assumption \ref{ass3}. Given the current time $t$, observations $X_{\text{obs}}$ from $X\sim \mathcal{D}$, and a failure probability $\delta\in(0,1)$, compute a lower bound  $\rho^*$ such that $\text{Prob}(\rho^\phi(X,\tau_0)\ge \rho^*)\ge 1-\delta$.  
\end{problem} 

Once we have computed the lower bound $\rho^*$, we remark that $\text{Prob}\big((X,\tau_0)\models \phi\big)\ge 1-\delta$ if $\rho^*>0$, see \cite[Proposition 16]{fainekos2009robustness}.   We next introduce a running example that we use throughout this paper to explain and illustrate our algorithms.


\begin{example}
\label{example:1}
We consider the F-16 aircraft  from \cite{heidlauf2018verification} with a hybrid controller modelled by $16$ states. We  only consider the height $h$ (given in ft) as the state to verify the specification $\phi:=G_{[0, 105]} h \ge 60$ and to find $\rho^*$ from Problem \ref{prob1} for $\delta:=0.2$. To construct a simple academic example, we collect a single trajectory $x_c$ from the simulator and then add independent noise to $x_c$ at each time\footnote{We note that it is possible to add noise directly to the dynamics of the F-16 aircraft and collect trajectory data from there.}, i.e., we let $\mathcal{N}(x_c(t), 3^2)$ and $\mathcal{N}(x_c(t), 3.5^2)$ describe $\mathcal{D}_0$ and $\mathcal{D}$, respectively. We assume that we have $K:=2000$ calibration trajectories $X^{(i)}$ with $i\in\{1,\hdots,K\}$ from $\mathcal{D}_0$ as per Assumption \ref{ass1}.  
\end{example}



\section{Robust Conformal Prediction}
\label{sec:intro_conf}

Our solution to Problem \ref{prob1} will rely on trajectory predictors that predict future system states $X_\text{un}$ from observations $X_\text{obs}$. To quantify the accuracy of these predictions, we use the calibration dataset $S$ from $\mathcal{D}_0$ along with robust conformal prediction as presented in \cite{cauchois2020robust} to account for the distribution shift between $\mathcal{D}_0$ and $\mathcal{D}$. Robust conformal prediction is an extension of conformal prediction which is a  statistical tool for uncertainty quantification \cite{vovk2005algorithmic,shafer2008tutorial, angelopoulos2021gentle,lei2018distribution}.
\\[12pt]
\noindent\textbf{Conformal Prediction (CP).} Let $R^{(0)},\hdots,R^{(K)}\sim \mathcal{R}_0$ be $K+1$ independent and identically distributed random variables following a training distribution $\mathcal{R}_0$.\footnote{In fact, exchangeability of $R^{(0)},\hdots,R^{(K)}$ would be sufficient which is a weaker requirement than being independent and identically distributed.} The variable $R^{(i)}$ can be freely defined and is referred to as the nonconformity score. In regression, a common choice for $R^{(i)}$ is the prediction error $|Z^{(i)}-\mu(U^{(i)})|$ where the predictor $\mu$ attempts to predict   $Z^{(i)}$ based on an input $U^{(i)}$. We note that a large nonconformity score indicates a large prediction error.  Our goal is thus to obtain an upper bound for $R^{(0)}$ (our test data) from $R^{(1)},\hdots,R^{(K)}$ (our calibration data).  Formally, given a failure probability $\delta\in (0,1)$, we want to compute a constant $C$ (which depends on $R^{(1)},\hdots,R^{(K)}$) such that 
\setlength{\abovedisplayskip}{2pt}
\setlength{\belowdisplayskip}{2pt}
\begin{align*}
\text{Prob}(R^{(0)}\le C)\ge 1-\delta.
\end{align*}

The probability $\text{Prob}(\cdot)$ is defined over the product measure of $R^{(0)},\hdots,R^{(K)}$.
By a simple statistical argument, one can obtain $C$ to be the $1/K$ corrected $(1-\delta)$th quantile of the empirical distribution of the values $R^{(1)},\hdots,R^{(K)}$, i.e., 
\setlength{\abovedisplayskip}{2pt}
\setlength{\belowdisplayskip}{2pt}
\begin{align}\label{eq:vanilla_quantile}
C:=\text{Quantile}_{(1+1/K)(1-\delta)}(R^{(1)},\hdots,R^{(K)}).
\end{align}
Formally, for $\beta\in [0,1]$,  the quantile function is defined as 
\begin{align*}
    \text{Quantile}_{\beta}(R^{(1)}, \dots, R^{(K)}):=\text{inf}\{z\in \mathbb{R}|\text{Prob}(Z\le z)\ge \beta\}
\end{align*} 
where the random variable $Z:=\sum_i \delta_{R^{(i)}}/K$ where $\delta_{R^{(i)}}$ is a dirac distribution centered at $R^{(i)}$. Equation \eqref{eq:vanilla_quantile} thus requires  $0\le (1+1/K)(1-\delta)\le 1$ and imposes the implicit lower bound $(K+1)(1-\delta) \le K$ on the  number of data $K$.  If this bound is satisfied and $R^{(1)},\hdots,R^{(K)}$ are sorted in non-decreasing order, we obtain $C:=R^{(p)}$ with $p:=\lceil (K+1)(1-\delta)\rceil$, i.e., $C$ is the $p$th smallest nonconformity score. We remark that we trivially have $C := \infty$ if $(K+1)(1-\delta) > K$.
\\[12pt]
\noindent\textbf{Robust CP.} In this paper, our test data is different from the training data. We thus use a robust version of conformal prediction based on \cite{cauchois2020robust}. Assume that $R^{(0)},\hdots,R^{(K)}$ are again independent, but not identically distributed in the sense that $R^{(0)}\sim\mathcal{R}$ while $R^{(1)},\hdots,R^{(K)}\sim\mathcal{R}_0$ where $\mathcal{R}$ is a test distribution. Under the assumption that test and training distributions $\mathcal{R}$ and $\mathcal{R}_0$ are close, the calibration data from the training distribution can still be used to bound $R^{(0)}$.

\begin{lemma}[Corollary 2.2 in \cite{cauchois2020robust}] \label{lemma:1}
    Let $R^{(0)},\hdots,R^{(K)}$ be independent random variables with $R^{(0)}\sim\mathcal{R}$ and $R^{(1)},\hdots,R^{(K)}\sim\mathcal{R}_0$ where the distributions $\mathcal{R}$ and $\mathcal{R}_0$ are such that $D_f(\mathcal{R},\mathcal{R}_0)\le \epsilon$. For a failure probability $\delta\in (0,1)$, it holds that
    \begin{align}\label{eq:guarantee_robust}
        \text{Prob}(R^{(0)}\le \tilde{C})\ge 1-\delta
    \end{align}
    where the constant $\tilde{C}$ is computed as
    \begin{align}\label{eq:C_tilde}
        \tilde{C}:= \text{Quantile}_{1-\tilde{\delta}}(R^{(1)},\hdots,R^{(K)})
    \end{align}
    where the adjusted confidence level $\tilde{\delta}$ is defined as
    \begin{align*}
        \tilde{\delta}&:=1-g^{-1}(1-\delta_n)
    \end{align*}
    and obtained by solving a series of convex optimization problems as
    \begin{align*}
        \delta_n&:=1-g\big((1+1/K)g^{-1}(1-\delta)\big),\\
        g(\beta)&:=\inf \{z\in[0,1]|\beta f(z/\beta)+ (1-\beta)f((1-z)/(1-\beta)) \le \epsilon\},\\
        g^{-1}(\tau)&:=\sup \{\beta\in[0,1]|g(\beta)\le \tau\}.
    \end{align*}
\end{lemma}

We remark that equation \eqref{eq:C_tilde} requires $0\le 1-\tilde{\delta}\le 1$ and poses restrictions on the number of data $K$, the failure probability $\delta$, and the distribution shift $\epsilon$ as we elaborate on later in the paper. Note further that $g$ and $g^{-1}$ are both solutions to convex programs. The solution to $g^{-1}(\tau)$ can thus be computed efficiently, e.g., using line search over $\beta\in(0,1)$. In some special cases, we can even obtain closed-form solutions for $g$, e.g., for $f(r):= \frac{1}{2}|r-1|$, associated with the total variation distance, we obtain $g(\beta)=\max(0,\beta-\epsilon)$.
\\[12pt]
\noindent\textbf{Induced Distribution Shift.} Note that in runtime verification we assume an $\epsilon$-bounded distribution shift in terms of the $f$-divergence on the trajectory level, as described by $\mathcal{D}$ and $\mathcal{D}_0$. In both runtime verification algorithms, it will be necessary to quantify the induced distribution shift of functions that are defined over $X\sim \mathcal{D}$ and $X_0\sim \mathcal{D}_0$. The following result follows trivially by the 
the data processing inequality.
\begin{lemma}[Data processing inequality]\label{lemma:2}
Let $\mathcal{D}$ and $\mathcal{D}_0$ be distributions such that $D_f(\mathcal{D},\mathcal{D}_0)\le \epsilon$ and let $R:\mathcal{X}\to \mathbb{R}$ be a measurable function. For $X\sim\mathcal{D}$ and $X_0\sim\mathcal{D}_0$, let $\mathcal{R}$ and $\mathcal{R}_0$ denote the induced distributions of $R(X)$ and $R(X_0)$, respectively. Then, it holds that
\setlength{\abovedisplayskip}{2pt}
\setlength{\belowdisplayskip}{2pt}
\begin{align*}
    D_f(\mathcal{D},\mathcal{D}_0)\le \epsilon \;\;\; \Rightarrow \;\;\; D_f(\mathcal{R},\mathcal{R}_0)\le \epsilon.
\end{align*}
\end{lemma}

\section{Predictive Runtime Verification under Distribution Shift}
\label{sec:f_div}

We propose two robust predictive runtime verification algorithms that account for the test distribution $\mathcal{D}$ being different from the training distribution $\mathcal{D}_0$. Our algorithms are inspired by our prior work in \cite{lindemann2023conformal}, but are able to deal with bounded distribution shifts $D_f(\mathcal{D},\mathcal{D}_0)\le \epsilon$. The first algorithm directly uses robust conformal prediction from Lemma \ref{lemma:1} to obtain a probabilistic lower bound $\rho^*$ for the robust semantics $\rho^\phi(X,\tau_0)$. This direct algorithm will provide a non-conservative verification result, but is lacking interpretability, i.e., if $\phi$ is violated no explanation is provided. The second algorithm therefore uses robust conformal prediction from Lemma \ref{lemma:1} to obtain a probabilistic lower bound for the robust semantics $\rho^\pi(X,\tau)$ of each predicate $\pi$ at each time $\tau$, which are subsequently used to obtain a probabilistic lower bound $\rho^*$ for $\rho^{\phi}(X,\tau_0)$. Besides being able to deal with distribution shifts, we remark that the indirect algorithm that we propose has significantly less conservatism compared to \cite{lindemann2023conformal}.
\\[12pt]
\noindent\textbf{Trajectory Predictor.} Our algorithms use trajectory predictors $\mu$ that map observations $X_\text{obs}$ at time $t$ into predictions $\hat{X}_{t + 1 | t} , \hat{X}_{t + 2 | t} \hdots$ of future states $X_\text{un}$. Therefore, we train a trajectory predictor $\mu$ on an additional training dataset from $\mathcal{D}_0$ that is independent of the calibration data in $S$. Commonly used trajectory predictors range from recurrent neural networks (RNN) and long short-term memory (LSTM) networks \cite{hochreiter1997long, fjellstrom2022long} to support vector machines \cite{cortes1995support, ristanoski2013time} and autoregressive integrated moving average models \cite{box2015time, mehrmolaei2016time}.

Since we consider bounded STL formulas $\phi$ as per Assumption \ref{ass3}, only a finite prediction horizon is needed. Therefore, we define $H := \tau_0 + L^\phi - t$ as the prediction horizon needed for the computation of satisfaction of $\phi$ imposed at $\tau_0$. To facilitate our discussion, we define the predicted trajectory 
\begin{align*}
    \hat{X} := (X_\text{obs}, \hat{X}_{t + 1 | t} , \hdots, \hat{X}_{t + H | t})
\end{align*}
with predictions $(\hat{X}_{t + 1 | t} , \hdots, \hat{X}_{t + H | t}):=\mu(X_\text{obs})$. We use the same notation for trajectories $X^{(i)}:=(X_\text{obs}^{(i)},X_\text{un}^{(i)})$ from the calibration dataset $S$. Formally, we also define the predicted calibration trajectory $\hat{X}^{(i)} := (X_\text{obs}^{(i)}, \hat{X}^{(i)}_{t + 1 | t} , \hdots, \hat{X}^{(i)}_{t + H | t})$ where $(\hat{X}^{(i)}_{t + 1 | t} , \hdots, \hat{X}^{(i)}_{t + H | t}):=\mu(X_\text{obs}^{(i)})$.

\subsection{Direct Robust Predictive Runtime Verification}
\label{subsec:robust_direct}
We first apply robust conformal prediction directly to the robust semantics $\rho^\phi$. To do so, we need to account for prediction errors in $\hat{X}$ and the distribution shift between the calibration trajectories $X^{(i)}\sim \mathcal{D}_0$ and the test trajectory $X\sim \mathcal{D}$. Specifically, consider the nonconformity score
\setlength{\abovedisplayskip}{2pt}
\setlength{\belowdisplayskip}{2pt}
\begin{equation}
\label{eq:R_direct}
    R^{(i)} = \rho^\phi(\hat{X}^{(i)}, \tau_0) - \rho^\phi(X^{(i)}, \tau_0)
\end{equation}
\noindent
for each calibration trajectory $X^{(i)}\in S$. Intuitively, this nonconformity score measures the difference between the predicted robust semantics $\rho^\phi(\hat{X}^{(i)}, \tau_0)$ and the true robust semantics $\rho^\phi(X^{(i)}, \tau_0)$. For the test trajectory $X$, we analogously define the test nonconformity score 
\setlength{\abovedisplayskip}{2pt}
\setlength{\belowdisplayskip}{2pt}
\begin{align*}
    R := \rho^\phi(\hat{X}, \tau_0) - \rho^\phi(X, \tau_0)
\end{align*}
\noindent
which we naturally cannot compute during runtime as $X$ is unknown. We denote the induced distribution of $R^{(i)}$ and $R$ by $\mathcal{R}_0$ and $\mathcal{R}$, respectively, so that $R^{(i)} \sim \mathcal{R}_0$ and $R \sim \mathcal{R}$, i.e., $\mathcal{R}$ is a shifted version of the distribution $\mathcal{R}_0$.

\begin{figure*}
    \centering
    \begin{subfigure}[t]{0.3\textwidth} 
        \includegraphics[width=\textwidth]{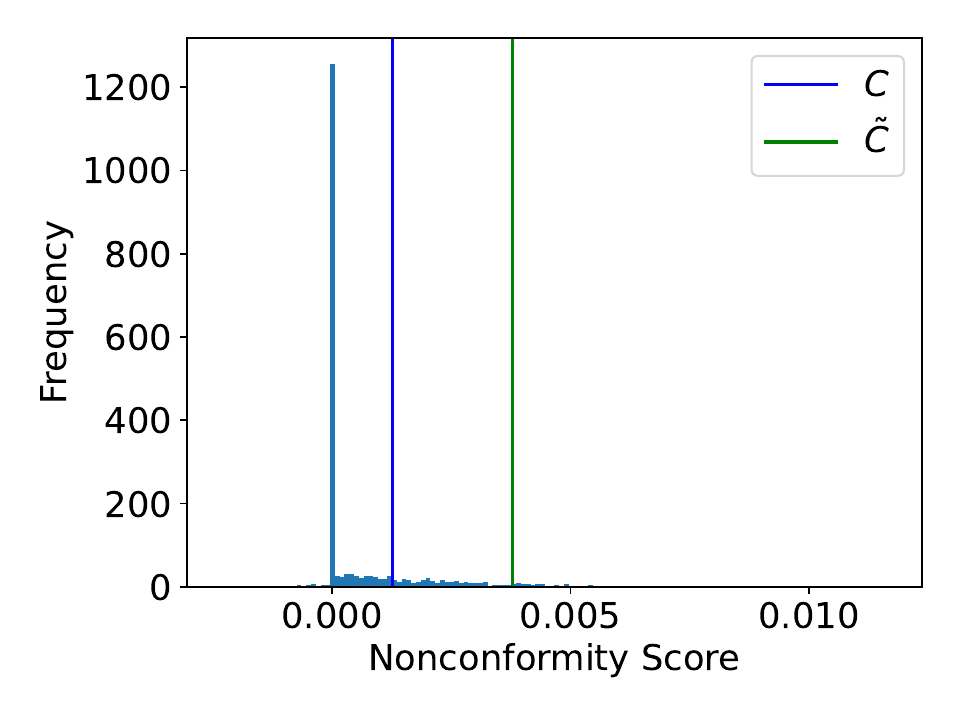}
        \caption{Histogram of $R^{(i)}$ from \eqref{eq:R_direct} over calibration data in $S$ along with robust prediction region $\tilde{C}$ from \eqref{eq:C_tilde} (and $C$ from \eqref{eq:vanilla_quantile}).}
        \label{fig:direct_nonconformities}
    \end{subfigure}
    \hfill
    \begin{subfigure}[t]{0.3\textwidth}
        \includegraphics[width=\textwidth]{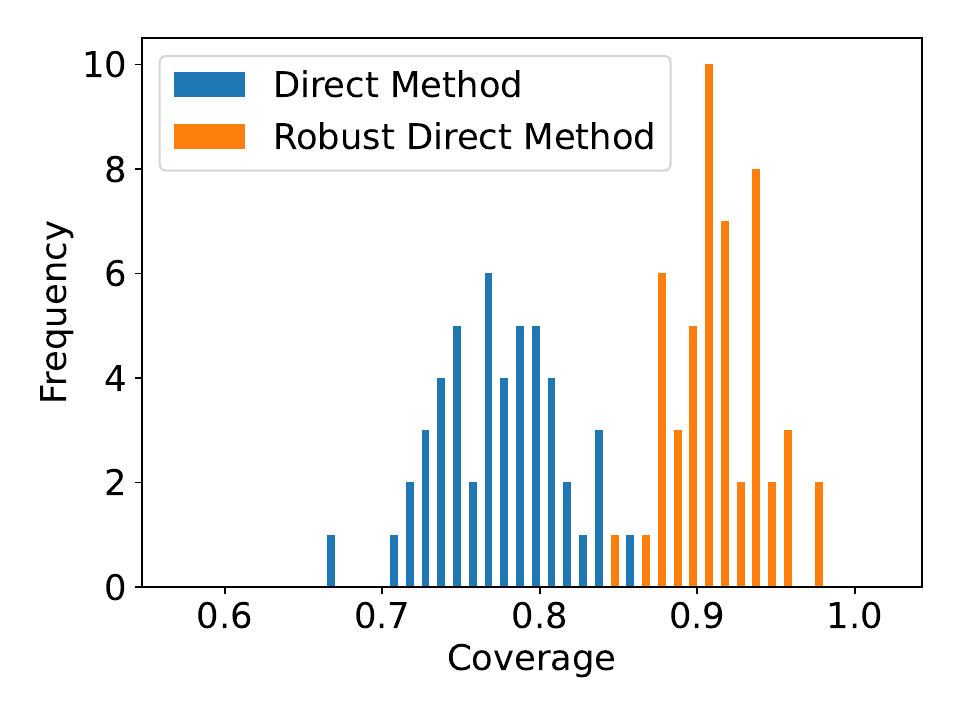}
        \caption{Histogram of empirical coverage on $\rho^\phi(X, \tau_0) \ge \rho^*$ for the robust direct and the non-robust direct method from \cite{lindemann2023conformal}.}
        \label{fig:direct_coverages}
    \end{subfigure}
    \hfill
    \begin{subfigure}[t]{0.3\textwidth} 
        \includegraphics[width=\textwidth]{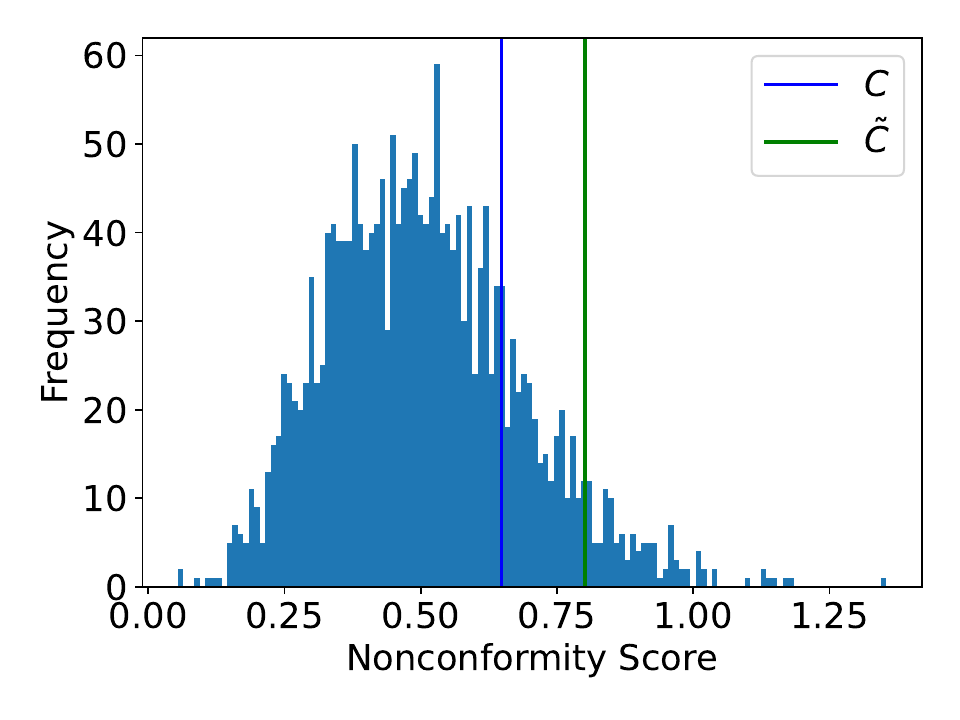}
        \caption{Histogram of $R^{(i)}$ from \eqref{eq:R_indirect} over calibration data in $S$ along with robust prediction region $\tilde{C}$ from \eqref{eq:C_tilde} (and $C$ from \eqref{eq:vanilla_quantile}).}
        \label{fig:indirect_1_nonconformities}
    \end{subfigure}
    \hfill
    \begin{subfigure}[t]{0.31\textwidth} 
        \includegraphics[width=\textwidth]{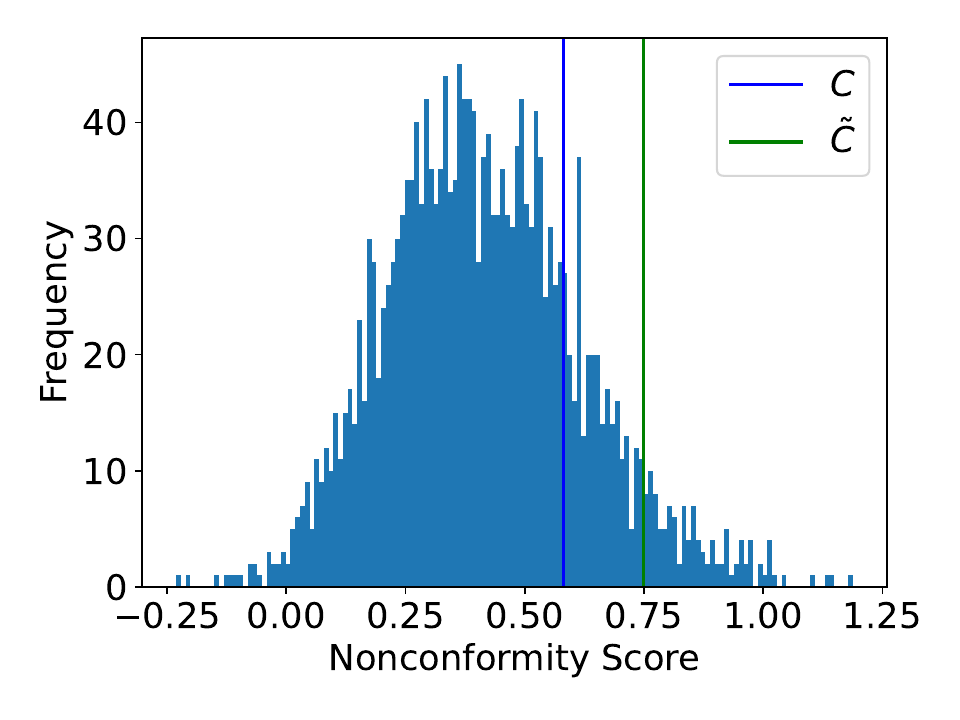}
        \caption{Histogram of $R^{(i)}$ from \eqref{eq:R_hybrid} over calibration data in $S$ along with robust prediction region $\tilde{C}$ from \eqref{eq:C_tilde} (and $C$ from \eqref{eq:vanilla_quantile}).}
        \label{fig:indirect_2_nonconformities}
    \end{subfigure}
    \hspace{5mm}
    \begin{subfigure}[t]{0.3\textwidth}
    \includegraphics[width=\textwidth]{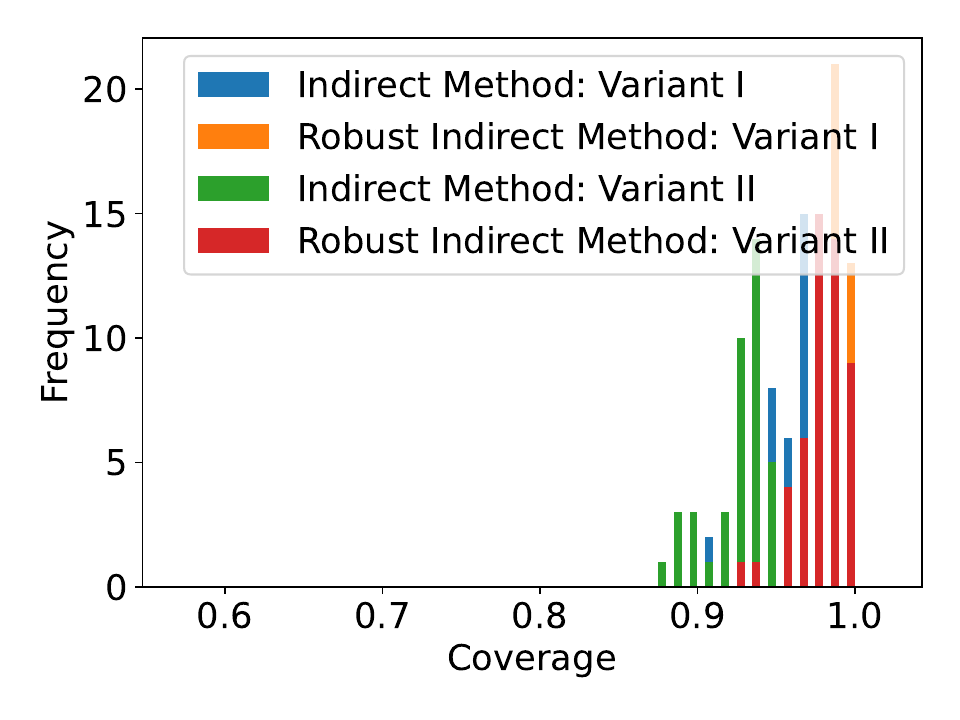}
    \caption{Histogram of empirical coverage on $\rho^\phi(X, \tau_0) \ge \rho^*$ with $C$ for the robust indirect methods (Variant I and II).}
    \label{fig:f_16_indirect_coverages}
    \end{subfigure}
    \caption{Running example: Histograms of nonconformity scores \eqref{eq:R_direct}, \eqref{eq:R_indirect}, and \eqref{eq:R_hybrid},  and empirical coverage plots of $\rho^\phi(X, \tau_0) \ge \rho^*$ for the direct and indirect (Variant I and II) methods.}
    \label{fig:combined}
\end{figure*}

By this construction of $R^{(i)}$ and $R$, it is easy to see from Lemmas \ref{lemma:1} and \ref{lemma:2} that $\text{Prob}((\rho^\phi(\hat{X}, \tau_0) - \rho^\phi(X, \tau_0) \le \tilde{C})\ge 1-\delta$ where $\tilde{C} := \text{Quantile}_{1 - \tilde{\delta}}(R^{(1)}, \hdots, R^{(K)})$ and $\tilde{\delta} := 1 - g^{-1}(1 - \delta_n)$. We summarize these results in Theorem \ref{theorem:1}, and provide a brief and formal proof in Appendix \ref{app:proof}.

\begin{theorem}\label{theorem:1}
Let the conditions from Problem \ref{prob1} hold. Then, it holds that $Prob(\rho^\phi(X, \tau_0) \ge \rho^{\phi}(\hat{X},\tau_0)-\tilde{C}) \ge 1 - \delta$ where $\tilde{C}$ is computed as in \eqref{eq:C_tilde} with the nonconformity score $R^{(i)}$ in \eqref{eq:R_direct} defined for all calibration trajectories $X^{(i)}\in S$.
\end{theorem}

Intuitively, this result says that the robust semantics $\rho^\phi(X, \tau_0)$ is lower bounded by the predicted robust semantics $\rho^\phi(\hat{X},\tau_0)$ adjusted by the value $\tilde{C}$, and that this holds with high probability. While the construction of $R^{(i)}$ and $R$ for the direct method follows \cite{lindemann2023conformal}, the result in Theorem \ref{theorem:1} accounts for deviations between training and test distributions and is not limited to the case where $\mathcal{D}$ is identical to $\mathcal{D}_0$. 
\begin{example}
\label{example:2}
    Recall Example \ref{example:1}. We set $t:=100$ and train an LSTM on $500$ trajectories from $\mathcal{D}_0$ to predict the next $H:=5$ time steps. We then perform the following experiment $50$ times: we sample $2000$ calibration trajectories from $\mathcal{D}_0$ and $100$ test trajectories from $\mathcal{D}$. For one of these experiments, we show in Figure \ref{fig:direct_nonconformities} the histogram of nonconformity scores $R^{(i)}$ from \eqref{eq:R_direct} along with the robust prediction region $\tilde{C}$ computed as in \eqref{eq:C_tilde}. In this case, we computed $\tilde{C}$ for the total variation distance with $\epsilon:=0.142$, which is such that $TV(\mathcal{R},\mathcal{R}_0)\le \epsilon$. In Section \ref{sec:simulations}, we explain in more detail how we estimate $TV(\mathcal{R},\mathcal{R}_0)$ in practice.  For comparison, we also plot the non-robust prediction region $C$ from \cite{lindemann2023conformal} which corresponds to the case where $\epsilon=0$ and which  is smaller than $\tilde{C}$  as it does not account for distribution shifts. This becomes evident in Figure \ref{fig:direct_coverages} where we plot the empirical coverage over all $50$ experiments for both algorithms. In other words, for each experiment we compute the ratio of how many of the $100$ test trajectories satisfy $\rho^\phi(X^{(i)}, \tau_0) \ge \rho^{\phi}(\hat{X}^{(i)},\tau_0)-\tilde{C}$ (where $\tilde{C}$ is replaced by $C$ for the non-robust version from \cite{lindemann2023conformal}), and then plot the histogram over these ratios. As we aim for $1-\delta=0.8$ coverage, we can observe that only the robust algorithm from Theorem \ref{theorem:1} achieves the desired coverage.
\end{example}

\subsection{Interpretable Robust Predictive Runtime Verification}
\label{subsec:robust_indirect}
This direct algorithm provides a precise verification result, but lacks interpretability when the specification is violated, i.e., no information for the cause of violation is provided. Therefore, assume that the formula $\phi$ is in positive normal form, i.e., that $\phi$ contains no negations. Note that every  $\phi$ can be re-written in positive normal form, see e.g., \cite{sadraddini2015robust}. Further, let the formula $\phi$ consists of $m$ predicates $\pi_i$, and define  
\begin{align*}
    \mathcal{P} := \big\{(\pi_i, \tau)| i\in\{1,\hdots,m\}, \tau\in \{t+1,\hdots,t+H\}\big\}
\end{align*} 
as the set of all predicates and times. We now design an interpretable robust predictive runtime verification algorithm, which we refer to as the robust indirect method, that uses probabilistic lower bounds $\rho_{\pi,\tau}^*$ of the robust semantics $\rho^{\pi}(X,\tau)$ for all predicates and times $(\pi,\tau)\in \mathcal{P}$, i.e., such that
\begin{align}\label{eq:guarantee_prediate}
\text{Prob}(\rho^{\pi}(X, \tau) \ge \rho_{\pi,\tau}^*, \forall (\pi, \tau) \in \mathcal{P}) \ge 1 - \delta.
\end{align}

Intuitively, $\rho_{\pi,\tau}^*\ge 0$ certifies that $\pi$ is satisfied at time $\tau$ with probability $1-\delta$. We thus obtain interpretable information via the lower bounds $\rho_{\pi,\tau}^*$ in the presence of distribution shifts. Before we propose two ways of computing $\rho_{\pi,\tau}^*$ from the predicted trajectory $\hat{X}$, we state our main results upfront. Therefore, we recursively define the probabilistic robust semantics $\bar{\rho}^\phi$, which provide the desired probabilistic lower bound $\rho^*$ in Problem \ref{prob1}, starting from predicates as
\begin{align*}
     \bar{\rho}^{\pi}(\hat{X},\tau)& := 
     \begin{cases}
     h(X_\tau) &\text{ if } \tau\le t\\
     \rho_{\pi,\tau}^* &\text{ otherwise }
     \end{cases}
 \end{align*} 
 while the other Boolean and temporal operators follow standard semantics, as summarized in Appendix \ref{app:STL}. We remark that the construction of $\bar{\rho}^{\pi}$ is again inspired by \cite{lindemann2023conformal}, but that we generalize and provide two ways to compute bounds $\rho_{\pi,\tau}^*$ that are i) robust to distribution shift, and ii) less conservative as we elaborate on more below.  We next state our main results for which we provide a proof in Appendix \ref{app:proof}.
\begin{theorem}\label{theorem:2}
Let the conditions from Problem 1 hold, and let $\phi$ further be in positive normal form. If the lower bounds $\rho^*_{\pi, \tau}$  satisfy  equation \eqref{eq:guarantee_prediate}, then it holds that  $\text{Prob}(\rho^{\phi}(X, \tau_0) \ge \bar{\rho}^\phi(\hat{X},\tau_0)) \ge 1 - \delta$ where $\bar{\rho}^\phi(\hat{X},\tau_0)$ is recursively constructed from $\rho^*_{\pi, \tau}$ as previously described.
\end{theorem}
\noindent \textbf{Computing $\rho_{\pi,\tau}^*$ on the state level (Variant I).} We now present two ways to compute $\rho_{\pi,\tau}^*$ that satisfy equation \eqref{eq:guarantee_prediate}. In the first method (called Variant I), we compute prediction regions for trajectory predictions via robust conformal prediction. Therefore, we define the nonconformity score 
\begin{align}\label{eq:R_indirect}
    R^{(i)} := \max_{\tau \in \{t+1, \hdots, t+H\}} \frac{\|X^{(i)}_\tau - \hat{X}^{(i)}_{\tau|t}\|}{\alpha_\tau}
\end{align}
where $\alpha_\tau>0$ are constants that normalize the prediction errors at times $\tau$, following a similar idea to \cite{cleaveland2023lcp}. In this work, however, we simply propose to compute $\alpha_\tau:=\max_i \|X^{(i)}_\tau - \hat{X}^{(i)}_{\tau|t}\|$ over an additional set of trajectories $X^{(i)}$ from $\mathcal{D}_0$ that is independent from the dataset $S$, such as the set of training trajectories used to train the predictor $\mu$ on. Next, we define $\mathcal{B}_\tau := \{\zeta\in\mathbb{R}^n | \|\zeta - \hat{X}_{\tau|t}\| \le \tilde{C}\alpha_\tau\}$ which is a norm ball of radius $\tilde{C}\alpha_\tau$ with center at $\hat{X}_{\tau|t}$. We then compute the worst case value of $\rho^\phi(\zeta,\tau)$ over all $\zeta\in \mathcal{B}_{\tau}$, i.e., we let 
\begin{align}\label{eq:worst_case_robustness}
    \rho^*_{\pi, \tau} = \inf_{\zeta\in \mathcal{B}_{\tau}} h(\zeta).
\end{align}

Finally, we relate this construction to equation \eqref{eq:guarantee_prediate} and provide a proof for the following result in Appendix \ref{app:proof}.

\begin{lemma}\label{lemma:variant1}
    Let the conditions from Problem \ref{prob1} hold. If $\alpha_\tau>0$ for all $\tau \in \{t+1, \hdots, t+H\}$, then it holds that 
    \begin{align} \label{eq:prob_indirect_1}
    \text{Prob}(\|X_\tau - \hat{X}_{\tau|t}\|\le \tilde{C}\alpha_\tau,  \forall \tau \in \{t+1, \hdots, t+H\}) \ge 1 - \delta
\end{align}
where $\tilde{C}$ is computed as in \eqref{eq:C_tilde} with the nonconformity score $R^{(i)}$ in \eqref{eq:R_indirect} defined for all calibration trajectories $X^{(i)}\in S$. Under the same conditions, it holds that $\rho^*_{\pi, \tau}$ in \eqref{eq:worst_case_robustness} satisfy~\eqref{eq:guarantee_prediate}.
\end{lemma}

Theorem \ref{theorem:2} and Lemma \ref{lemma:variant1} together present an interpretable predictive runtime monitor that can account for distribution shifts between $\mathcal{D}_0$ and $\mathcal{D}$ via the values of $\rho^*_{\pi, \tau}$. Finally, we emphasize that the normalization via the constants $\alpha_\tau$ greatly reduces the conservatism presented in \cite{lindemann2023conformal} where the confidence level $1-\delta$ has to be adjusted to $1-\delta/H$.
\\[12pt]
\noindent\textbf{Computing $\rho_{\pi,\tau}^*$ on the predicate level (Variant II).}  While Variant I provides interpretability, it may be the case that taking the infimum in equation \eqref{eq:worst_case_robustness} is conservative, e.g., as we show later in Figure \ref{fig:case_study_indirect_robustnesses} in our case study. We thus present a second method (called Variant II) where we compute prediction regions for each predicate $\pi$ directly. Therefore, consider the nonconformity score
\begin{align}\label{eq:R_hybrid}
    R^{(i)} := \max_{(\pi, \tau) \in \mathcal{P}}\frac{\rho^{\pi}(\hat{X}^{(i)}, \tau) - \rho^{\pi}(X^{(i)}, \tau)}{\alpha_{\pi, \tau}}
\end{align}
where $\alpha_{\pi, \tau}>0$ are again normalization constants. In this work, we use $\alpha_{\pi, \tau}:=\max_i |\rho^{\pi}(\hat{X}^{(i)}, \tau) - \rho^{\pi}(X^{(i)}, \tau)|$ over an additional set of trajectories $X^{(i)}$ from $\mathcal{D}_0$ that is independent from $S$. We conclude with the following result for which we provide a proof in Appendix \ref{app:proof}.
\begin{lemma}\label{lemma:variant2}
Let the conditions from Problem \ref{prob1} hold. If $\alpha_{\pi, \tau}>0$ for all $\tau \in \{t+1, \hdots, t+H\}$, then it holds that 
    \begin{align} \label{eq:prob_indirect_2}
    \text{Prob}(\rho^{\pi}(\hat{X}, \tau) - \rho^{\pi}(X, \tau) \le \tilde{C}\alpha_{\pi, \tau}, \forall (\pi, \tau) \in \mathcal{P}) \ge 1 - \delta
\end{align}
where $\tilde{C}$ is computed as in \eqref{eq:C_tilde} with the nonconformity score $R^{(i)}$ in \eqref{eq:R_hybrid} defined for all calibration trajectories $X^{(i)}\in S$. Under the same conditions, it holds that 
\begin{align*}
    \rho^*_{\pi, \tau}:=\rho^{\pi}(\hat{X}, \tau)-\tilde{C}\alpha_{\pi, \tau}
\end{align*}
satisfy~\eqref{eq:guarantee_prediate}.
\end{lemma}

\begin{example}
    \label{example:3}
    Recall Example \ref{example:1}. Similar as for the direct algorithm in Example \ref{example:2}, we perform the same experiment of sampling $2000$ calibration trajectories and $100$ test trajectories $50$ times. For one of these experiments, we show in Figures \ref{fig:indirect_1_nonconformities} and \ref{fig:indirect_2_nonconformities} the histograms of the nonconformity scores $R^{(i)}$ from \eqref{eq:R_indirect} and \eqref{eq:R_hybrid} along with the robust prediction region $\tilde{C}$. As in Example \ref{example:2}, we use $\epsilon:=0.142$ which is such that $TV(\mathcal{R},\mathcal{R}_0)\le \epsilon$  where the induced distributions $\mathcal{R}$ and $\mathcal{R}_0$ are now with respect to equations \eqref{eq:R_indirect} and \eqref{eq:R_hybrid}. For comparison, we also plot the non-robust prediction region $C$ from \cite{lindemann2023conformal} which corresponds to the the case where $\epsilon=0$. In Figure \ref{fig:f_16_indirect_coverages},  we plot the histogram over all $50$ experiments of the empirical coverage for $\text{Prob}(\rho^{\phi}(X^{(i)}, \tau_0) \ge \bar{\rho}^\phi(\hat{X}^{(i)},\tau_0) \ge 1 - \delta$ on test trajectories $X^{(i)}$  for Variants I and II. We achieve the desired coverage of $1-\delta=0.8$ and compare to the non-robust versions (using $C$ instead of $\tilde{C}$). In this case, these also achieve an empirical coverage of $0.8$ as the indirect algorithms are more conservative than the direct algorithm.
\end{example}

\begin{figure*}
    \centering
    \begin{subfigure}[t]{0.4\textwidth}
        \includegraphics[width=\linewidth]{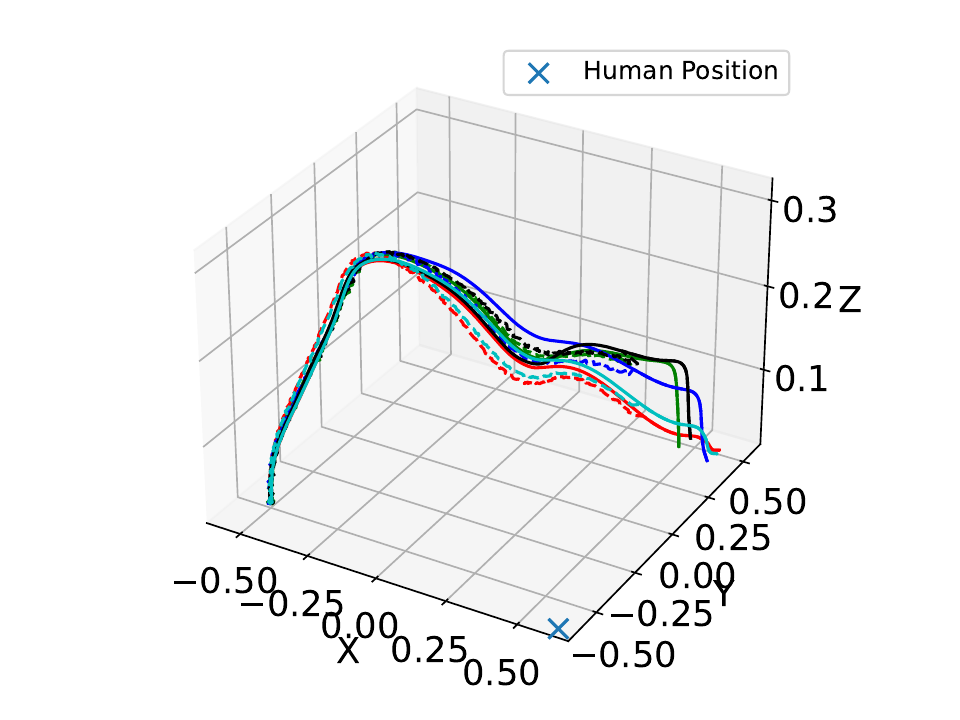}
        \caption{Trajectories from $\mathcal{D}_0$ with ground truth in solid and predictions in dashed lines for $t = 400$ and $H = 516$.}
         \label{fig:trajectories_original}
    \end{subfigure}
    \hspace{5mm}
     \begin{subfigure}[t]{0.4\textwidth}
        \includegraphics[width=\linewidth]{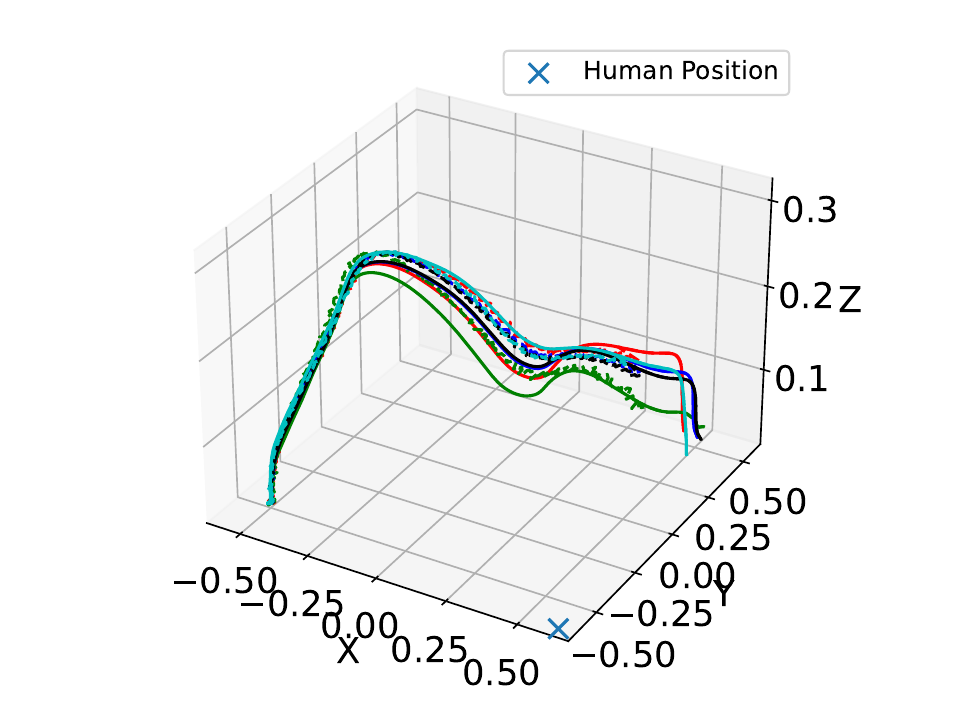}
        \caption{Trajectories from $\mathcal{D}$ with ground truth in solid and predictions in dashed lines when $t = 400$ and $H = 516$.}
         \label{fig:trajectories_disturbed}
    \end{subfigure}
    \caption{Trajectories from $\mathcal{D}_0$ and $\mathcal{D}$ for the Franka robotic arm case study.}
    \label{fig:case_study_setup}
\end{figure*}

\subsection{Data Requirements and Feasibility}
The algorithms presented in Theorems \ref{theorem:1} and \ref{theorem:2} (in combination with Lemmas \ref{lemma:variant1} and \ref{lemma:variant2}) provide runtime verification guarantees when the prediction region $\tilde{C}$, which is computed according to Lemma \ref{lemma:1}, is finite and not unbounded (i.e.,  $\tilde{C}=\infty$). For this to happen, we require that $1-\tilde{\delta}\in [0,1]$ which is equivalent to the condition $1-\delta_n=(1+1/K)g^{-1}(1-\delta)\in [0,1]$ as the function $g$ and its inverse $g^{-1}$ have domains $[0,1]$. We note that the lower bound $0\le (1+1/K)g^{-1}(1-\delta)$ is trivially satisfied for any $K>0$. The upper bound $(1+1/K)g^{-1}(1-\delta)\le 1$, on the other hand, poses a lower bound on the number $K$ of calibration trajectories as
\begin{align}\label{eq:K_lower}
    K \ge \Big\lceil \frac{g^{-1}(1 - \delta)}{1 - g^{-1}(1 - \delta)} \Big\rceil.
\end{align}
This condition can be seen as a requirement on the number of calibration data $K$ if $g^{-1}(1 - \delta)<1$. However, it is important to observe the case where $g^{-1}(1 - \delta)=1$. It is this condition that imposes additional conditions on the confidence $1-\delta$ and the distribution shift $\epsilon$ for a given $f$-divergence. For instance, for $f(t) = \frac{1}{2}|t - 1|$, associated with the total variation distance, we know that 
\begin{align*}
    g^{-1}(1 - \delta)=\text{argsup}_{\beta\in[0,1]} \max(0,\beta-\epsilon)\le 1-\delta
\end{align*}
which is equivalent to $1$ if $\epsilon\ge \delta$. More generally, the condition $g^{-1}(1 - \delta)=1$ will constrain the permissible distribution shift $\epsilon$ for a confidence of $1-\delta$. 
\begin{corollary}
    Let the conditions from Problem \ref{prob1} hold. If equation \eqref{eq:K_lower} is satisfied with $g^{-1}(1 - \delta)<1$, then the algorithms presented in Theorems \ref{theorem:1} and \ref{theorem:2} (in combination with Lemmas \ref{lemma:variant1} and \ref{lemma:variant2}) provide nontrivial (i.e., $\tilde{C}<\infty$) results.
\end{corollary}
The direct and the indirect RPRV algorithms consider the confidence level $1-\delta$. This is  opposed to the indirect method in \cite{lindemann2023conformal} where the confidence level for the indirect method (Variant I) has to be adjusted to $1-\delta/H$, which in this setting would impose the condition  $\epsilon < \delta/H$ for $f(t) = \frac{1}{2}|t - 1|$.





\begin{remark}
    \textbf{Robust Locally Adaptive Predictive Runtime Verification:} Observing that the presented RPRV algorithms produce constant prediction regions $\tilde{C}$ regardless of the observation $X_\text{obs}$, we remark that our RPRV algorithms can be generalized to be locally adaptive \cite[Section 2]{angelopoulos2021gentle}. In this case, $\tilde{C}$ adapts to the observation $X_\text{obs}$ and the confidence of the trajectory predictor $\mu$ via a transformation of the nonconformity score. Specifically, one can consider a locally adaptive nonconformity score $R_{adapt}^{(i)} := R^{(i)}/\omega(X_{obs}^{(i)})$
    where $R^{(i)}$ is as in \eqref{eq:R_direct}, \eqref{eq:R_indirect}, or \eqref{eq:R_hybrid}
    and where $\omega$ is a predictor that estimates  $|R^{(i)}|$ from $X_{obs}^{(i)}$ with a trivial assumption on $\omega$ generating a positive value. At test time, it then holds that
    \begin{align}\label{eq:local_adaptive_guarantee}
    \text{Prob}(R_{adapt} \le \tilde{C} \omega(X_{obs})) \ge 1-\delta.
    \end{align}   
where $R_{adapt} \sim \mathcal{R}$ is derived from the test trajectory $X\sim \mathcal{D}$ and where $\tilde{C}:= \text{Quantile}_{1-\tilde{\delta}}(R_{adapt}^{(1)},\hdots,R_{adapt}^{(K)})$. Intuitively, a larger value of $\omega(X_{obs})$ indicates more uncertainty from the predictor $\mu$, thus increasing the prediction region to accommodate for heteroskedascity within the test data.
\end{remark}

\section{Case Study: Franka Manipulator}\label{sec:simulations}
\begin{figure*}
     \centering
     \begin{subfigure}[t]{0.32\textwidth}
         \centering
          \includegraphics[width=\textwidth]{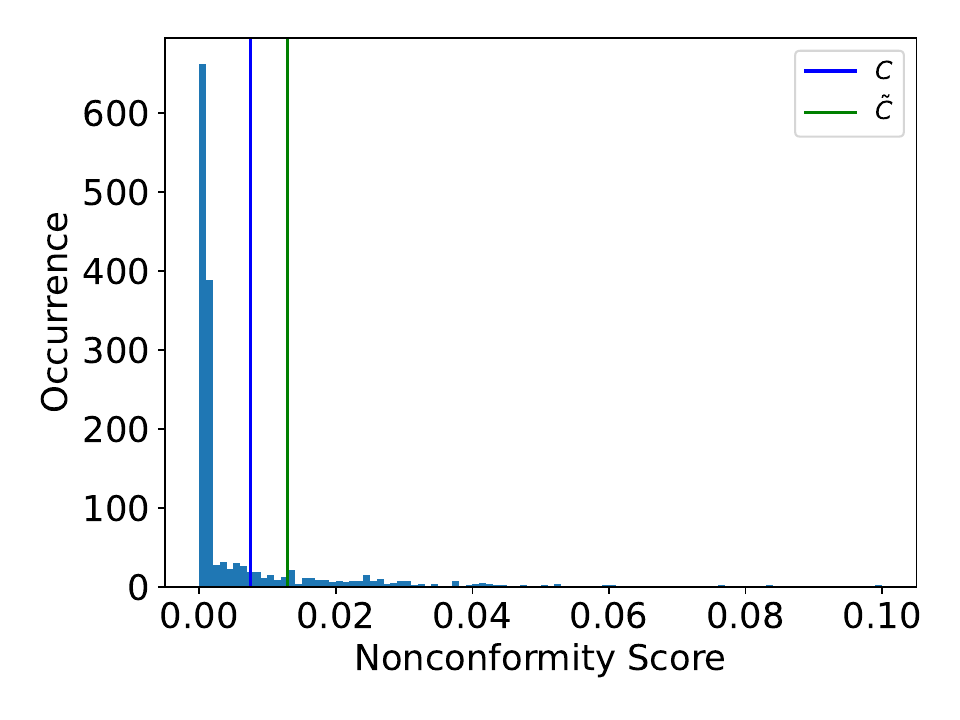}
          \caption{Histogram of $R^{(i)}$ from \eqref{eq:R_direct} over calibration data in $S$ along with robust prediction region $\tilde{C}$ from \eqref{eq:C_tilde} (and $C$ from \eqref{eq:vanilla_quantile}).}
    \label{fig:direct_nonconformity_distribution}
     \end{subfigure}
     \hfill
     \begin{subfigure}[t]{0.32\textwidth}
        \includegraphics[width=\textwidth]{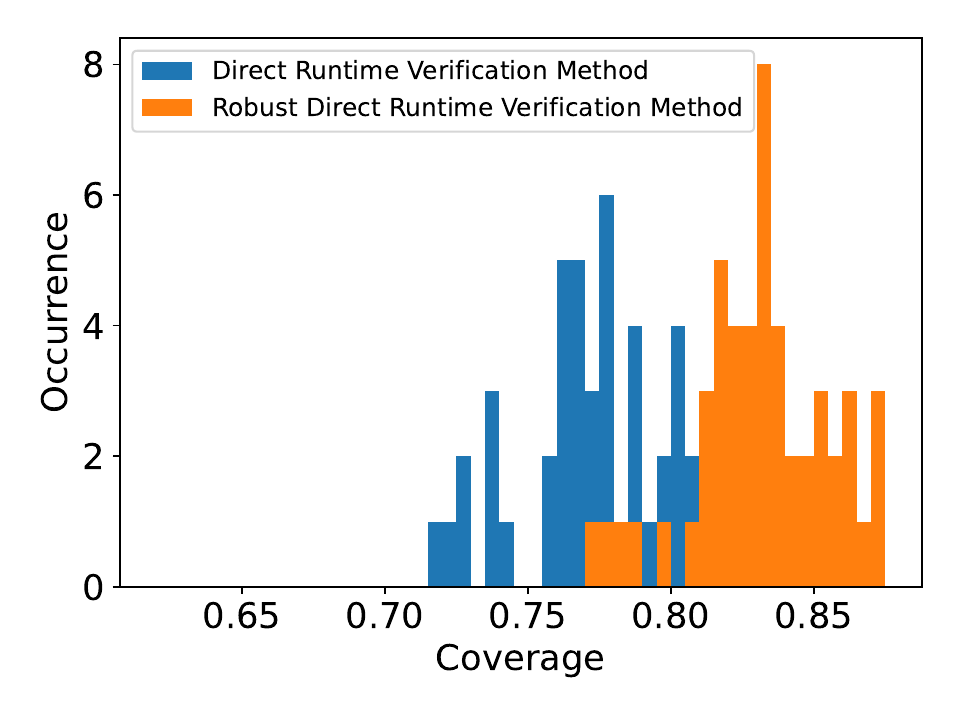}
        \caption{Histogram of empirical coverage on $\rho^\phi(X, \tau_0) \ge \rho^*$ for the robust direct and the non-robust direct method from \cite{lindemann2023conformal}.}
        \label{fig:case_study_direct_coverage}
     \end{subfigure}
     \hfill
     \begin{subfigure}[t]{0.32\textwidth}
        \includegraphics[width=\textwidth]{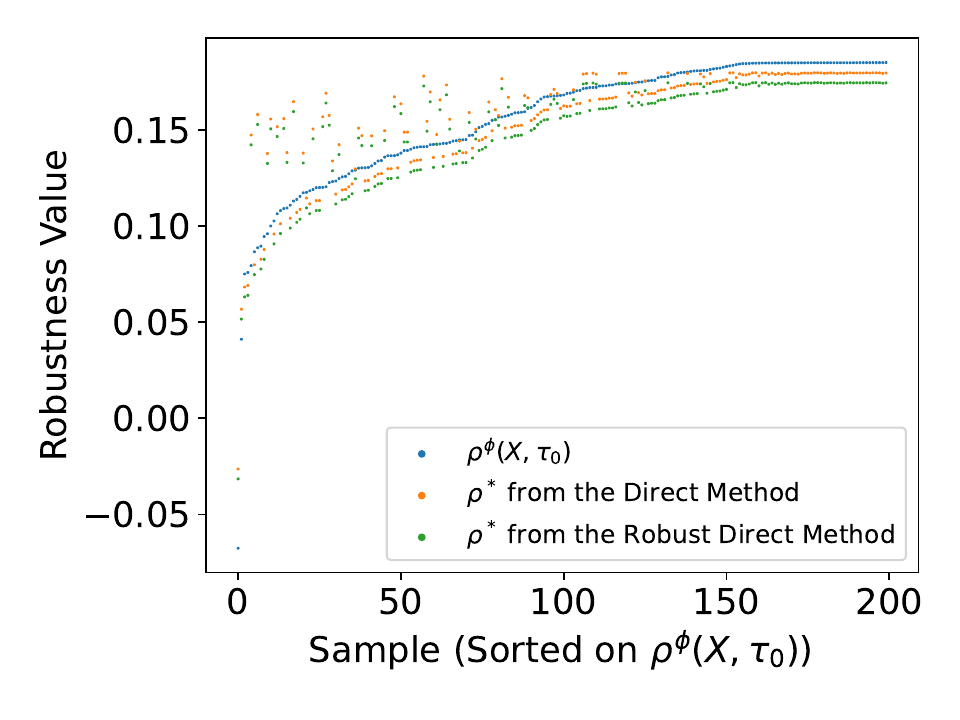}
        \caption{Ground truth robustness  $\rho^\phi(X, \tau_0)$ over a validation set in comparison with the predicted robustness $\rho^*$ for the robust direct and the non-robust direct method from \cite{lindemann2023conformal}.}
        \label{fig:case_study_direct_robustnesses}
     \end{subfigure}
     \caption{Case Study: Histograms of nonconformity scores, empirical coverage, and robust semantics plots for the direct method.}
     \label{fig:results_direct_methods}
\end{figure*}
To evaluate the proposed RPRV algorithms, we present a case study to verify the safety of a Franka robotic manipulator  within the NVIDIA Isaac Simulator.
\\[12pt]
\noindent \textbf{System Description:} The Franka, located at the origin of the workspace, is given the task to pick up a cubic box of length $0.03$ meters and place it at a target location $X^g := [-0.5, -0.5, 0]$. We use the Pick \& Place Controller from NVIDIA Isaac Sim \cite{makoviychuk2021isaac}. We also describe the state of the system  at time $t$ as $X_t := (X_t^p, X_t^s)$ where $X_t^p \in \mathbb{R}^3$ is the position of the box (controlled by the Franka robot) and $X_t^s$ is its linear speed (which is the norm of the linear velocity). We  describe the training and test distributions over trajectories $X_0^p$ as $\mathcal{D}_0:=[U(0.4, 0.6), U(0.4, 0.6), 0]$ and $\mathcal{D}:=[U(0.41, 0.61), U(0.41, 0.61), 0]$, respectively, where $U(a, b)$ denotes a uniform distribution over the range $[a,b]$. We only consider trajectories $X$ that successfully finish the placing task $\|X_{t+H}^p - X^g\|_2 < 0.3$ (i.e., we do not consider unsuccessful grasps). We assume to have access to a set of $1926$ trajectories from $\mathcal{D}_0$, which we denote by $Z_0$, for predictor training and calibration, and a set of $934$ trajectories from $\mathcal{D}$, denoted by $Z$,  for validating our algorithms. For illustration, we show five trajectories from $\mathcal{D}_0$ and five trajectories from $\mathcal{D}$ in solid lines along with the predictions in dashed lines in Figures \ref{fig:trajectories_original} and \ref{fig:trajectories_disturbed}, respectively.
\\[12pt]
\noindent \textbf{System Requirement:} The robot should not only place the box at the desired location, but the robot should in the process also present the box to a human inspector located at $X^h:=[0.6, -0.6, 0]$ with a moderate speed. Formally, we impose the following STL specification 
\begin{align*}
    \phi \models F_{[0, t + H]} (\|X_t^p - X_t^g\|_2 \le \zeta_1)  \wedge F_{[0, t + H]}(\|X_t^p - X^h\|_2 \\ \le \zeta_3) \wedge G_{[0, t + H]}((\|X_t^p - X^h\|_2 < \zeta_3) \implies (X_t^s \le \zeta_2))
\end{align*}
for  constants $\zeta_1 := 0.2$, $\zeta_2 := 0.6$, and $\zeta_3 := 0.9$. The formulas $F_{[0, t + H]} (\|X_t^p - X_g\|_2 \le \zeta_1)$ and $F_{[0, t+H]}(\|X_t^p - X^h\| \le \zeta_3)$ within $\phi$ are task completion requirements which specify that the robot successfully places the box at the goal location and in this process also presents the box to the inspector at $X^h$. The formula $ G_{[0, t+H]}((\|X_t^p - X^h\|_2 < \zeta_3) \implies (X_t^s \le \zeta_2))$ within $\phi$ uses the implication operator $\implies$ and expresses a safety requirement. Specifically, when the robot is close to the human inspector, the box should not exceed the speed limit of $\zeta_2$. As we require $\phi$ to be in positive normal form for the indirect methods, we emphasize that this last formula is equivalent to $G_{[0, t+H]}((\|X_t^p - X^h\|_2 >= \zeta_3) \vee (X_t^s \le \zeta_2))$, which we use in our implementation. 
\\[12pt]
\begin{figure*}
    \centering
    \begin{subfigure}[t]{0.4\textwidth}
        \includegraphics[width=\linewidth]{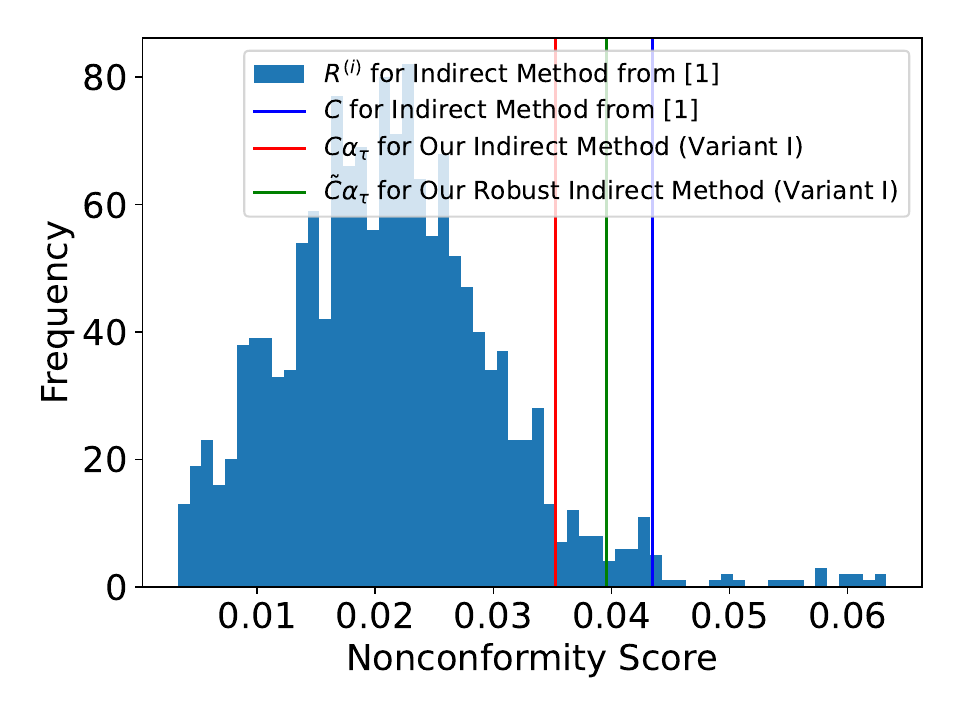}
        \caption{Histogram of $R^{(i)} = \|X^{(i)}_\tau - \hat{X}^{(i)}_\tau\|$ over calibration data in $S$ along with robust prediction region $\tilde{C}\alpha_\tau$ from Lemma \ref{lemma:variant1} (and $C\alpha_\tau$ with $C$ from \eqref{eq:vanilla_quantile}).}
         \label{fig:indirect_1_nonconformity_distribution}
    \end{subfigure}
    \hspace{5mm}
     \begin{subfigure}[t]{0.4\textwidth}
        \includegraphics[width=\linewidth]{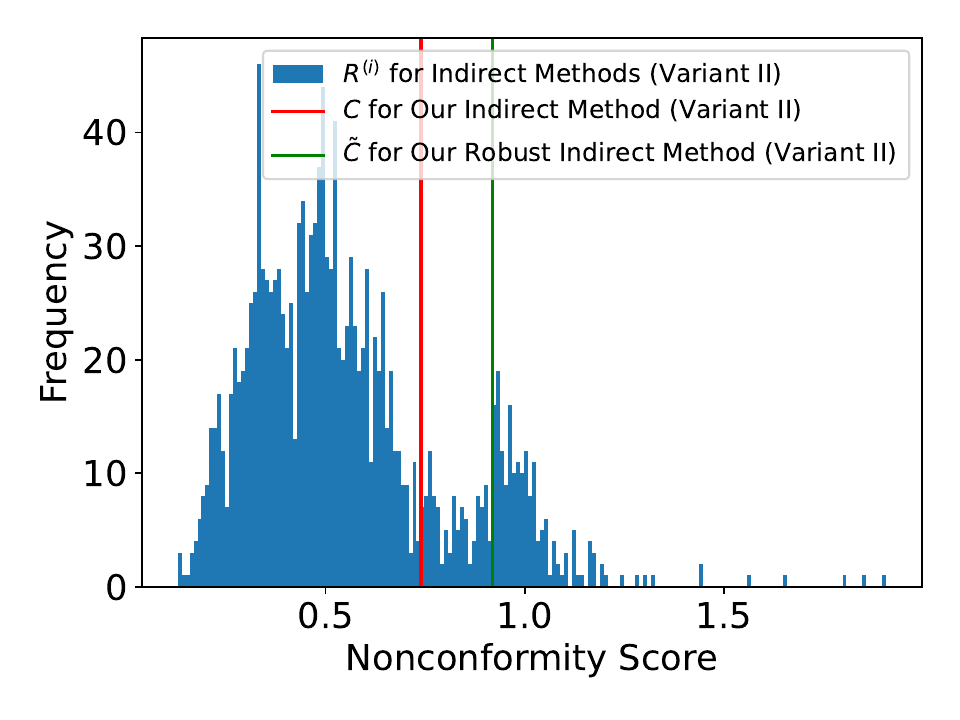}
        \caption{Histogram of $R^{(i)}$ from \eqref{eq:R_hybrid} over calibration data in $S$ along with robust prediction region $\tilde{C}$ from \eqref{eq:C_tilde} (and $C$ from \eqref{eq:vanilla_quantile}).}
         \label{fig:indirect_2_nonconformity_distribution}
    \end{subfigure}
    \begin{subfigure}[t]{0.4\textwidth}
        \includegraphics[width=\linewidth]{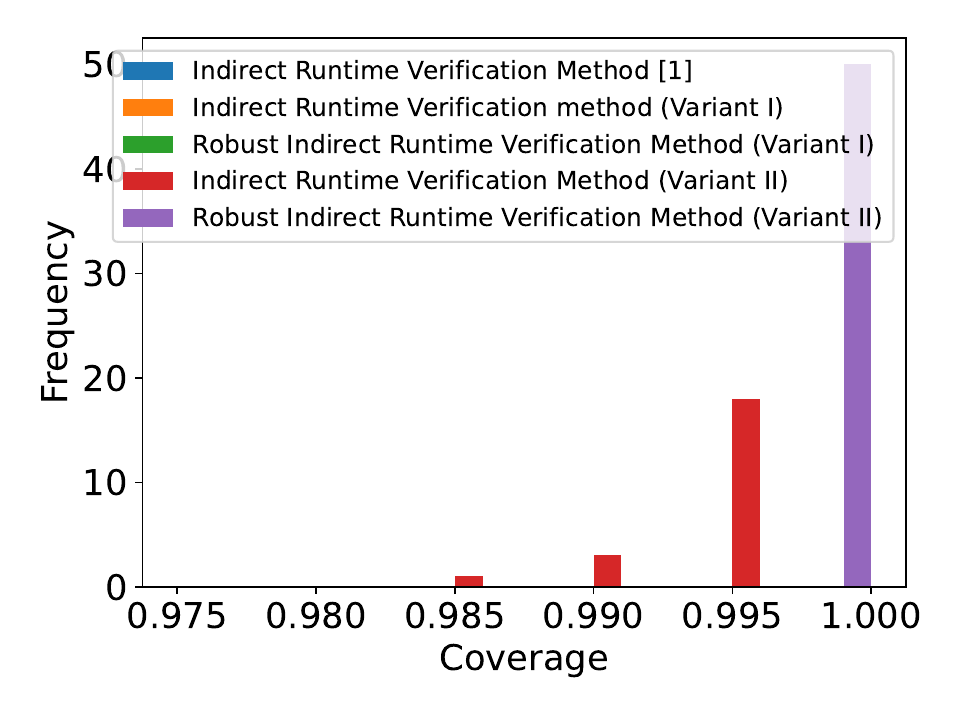}
        \caption{Histogram of empirical coverage on $\rho^\phi(X, \tau_0) \ge \rho^*$ for the robust indirect and the non-robust direct methods.}
         \label{fig:case_study_indirect_coverage}
    \end{subfigure}
    \hspace{5mm}
    \begin{subfigure}[t]{0.4\textwidth}
        \includegraphics[width=\linewidth]{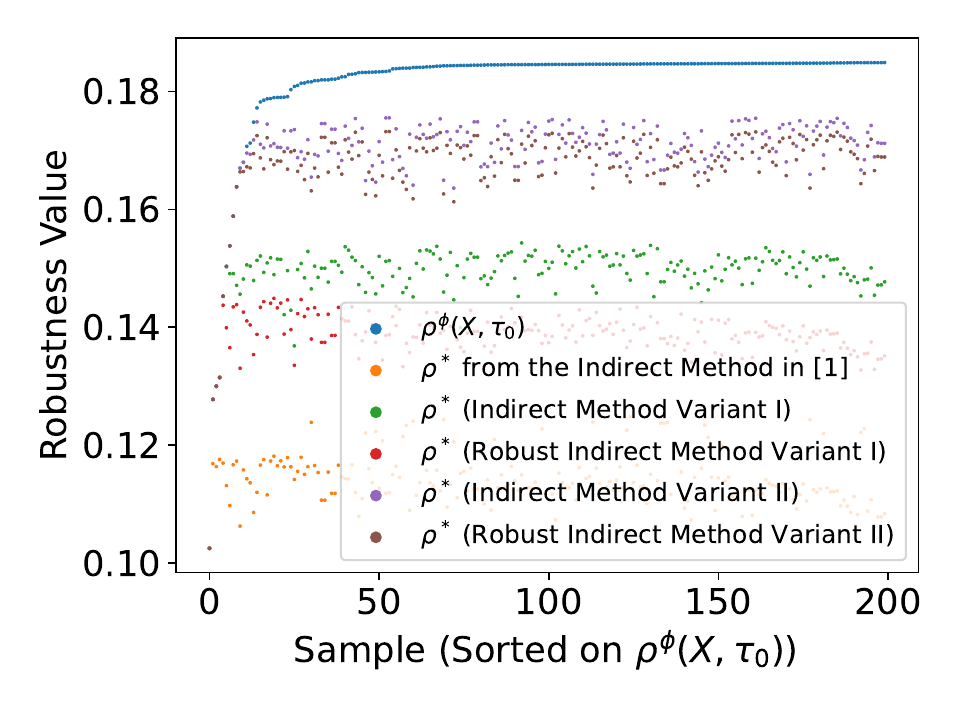}
        \caption{Ground truth robustness  $\rho^\phi(X, \tau_0)$ over a validation set in comparison with the predicted robustness $\rho^*$ for the robust indirect methods (variant I) and (variant II).}         \label{fig:case_study_indirect_robustnesses}
    \end{subfigure}
    \caption{Case Study: Histograms of nonconformity scores, empirical coverage, and robust semantics plots for the indirect methods.}
        \label{fig:nonconformity_indirect_methods}
\end{figure*}
\noindent\textbf{Distribution Shift Estimation:} For validation, we estimate the distribution shift $D_f( \mathcal{D}, \mathcal{D}_0)$. Therefore, we randomly pick 800 trajectories from $Z_0$ and $Z$. For each of the nonconformity scores $R^{(i)}$ from \eqref{eq:R_direct}, \eqref{eq:R_indirect}, and \eqref{eq:R_hybrid}, we perform the following procedure: we calculate $R^{(i)}$ for the trajectories sampled from $Z_0$ and $Z$ to obtain the empirical distributions of $\mathcal{R}_0$ and $\mathcal{R}$, respectively, where $\mathcal{R}_0$ and $\mathcal{R}$ are the induced distributions using the nonconformity scores over $\mathcal{D}$ and $\mathcal{D}_0$. Specifically, we use kernel density estimators with Gaussian kernels to estimate the empirical probability density functions (PDFs) of each distribution. We then  numerically evaluate the distribution shift by computing $TV(\mathcal{R}, \mathcal{R}_0) = \frac{1}{2}\int_x|q(x) - p(x)|dx$ where $p(x)$ and $q(x)$ are the estimated PDFs associated with $\mathcal{R}$ and $\mathcal{R}_0$. We obtain the values $\epsilon_1$, $\epsilon_2$, and $\epsilon_3$ that indicate the estimated distribution shifts on \eqref{eq:R_direct}, \eqref{eq:R_indirect}, and \eqref{eq:R_hybrid}. Finally, we take $\epsilon := \max(\epsilon_1, \epsilon_2, \epsilon_3)$ so that $\epsilon$ is greater than the estimated distribution shift of $ D_f(\mathcal{R}_0, \mathcal{R})$ for all $\mathcal{R}_0$ and $\mathcal{R}$ in \eqref{eq:R_direct}, \eqref{eq:R_indirect}, and \eqref{eq:R_hybrid}. Intuitively, this allows us to validate the statistical guarantees for all our verification algorithms. We note that we used the same method to calculate the distribution shift in the running example  (see Example \ref{example:1}), but with 1000 trajectories randomly generated from each distribution. We remark that $\epsilon$ can be thought of as a tuning parameter in practice, as it is common practice in robust control \cite{zhou1996robust}, and that the purpose of the estimation here is for validation of our algorithms.
\\[12pt]
\noindent \textbf{Validation and Comparison of Direct Method:} For this case study, we seek to find $\rho^*$ from Problem \ref{prob1} for a failure probability $\delta := 0.2$. We set $t := 400$ and predict the next $H := 516$ time steps via an LSTM trajectory predictor trained on $192$ trajectories sampled from $Z_0$. We run the following experiments $50$ times: We sample $K := 1500$ calibration trajectories from $Z_0$ but separate out the trajectories used for training the LSTM predictor. For one of these experiments, we show the histogram of nonconformity scores $R^{(i)}$ from \eqref{eq:R_direct} and the robust prediction region $\tilde{C}$ from \eqref{eq:C_tilde} in Figure \ref{fig:direct_nonconformity_distribution} with $\epsilon := 0.051$ based on the aforementioned method for distribution shift estimation. For comparison, we also show the non-robust prediction region $C$ from the direct method in \cite{lindemann2023conformal} (where $\epsilon = 0$), which is smaller than $\tilde{C}$ and cannot deal with the distribution shift. In Figure \ref{fig:case_study_direct_coverage}, we plot the empirical coverage over the $50$ experiments. Specifically, for each experiment, we sample a set of $200$ test trajectories from $Z$ and compute the percentage of trajectories  satisfying $\rho^\phi(X^{(i)}) \ge \rho^\phi(\hat{X}^{(i)}) - C$ and $\rho^\phi(X^{(i)}) \ge \rho^\phi(\hat{X}^{(i)}) - \tilde{C}$ for the non-robust and robust methods. For one experiment, we show the true robust semantics $\rho^\phi(X, \tau_0)$ for the 200 ground truth test data and the predicted worst-case robust semantics $\rho^*$ in Figure \ref{fig:case_study_direct_robustnesses} for the non-robust and robust RPRV algorithms.
\\[12pt]
\noindent \textbf{Validation and Comparison of Indirect Methods:} For practical reasons, we first downsample trajectories from 917 to 25 time steps in $Z_0$ and $Z$ to reduce the computational overhead from solving the optimization problem in \eqref{lemma:variant1} for Variant I. We set $t := 11$ and predict the next $H := 13$ time steps via an LSTM predictor trained on 192 trajectories from $Z_0$, and we again select $\delta := 0.2$. We perform the same validation experiment as for the direct method for $50$ times with $\epsilon := 0.076$, which we computed following the aforementioned method for estimating distribution shifts. \emph{Variant I.} For one of these experiments, we show the histogram of the nonconformity scores $R^{(i)} := \|X^{(i)}_\tau - \hat{X}^{(i)}_\tau\|$ from the indirect method in \cite{lindemann2023conformal} for time $\tau := 24$ in Figure \ref{fig:indirect_1_nonconformity_distribution}. For comparison, we also plot $C\alpha_\tau$ and $\tilde{C}\alpha_\tau$ following our result in Lemma \ref{lemma:variant1}  where $C$ and $\tilde{C}$ are computed from equations \eqref{eq:vanilla_quantile} and \eqref{eq:C_tilde} with $\alpha_\tau$ and $R^{(i)}$ in \eqref{eq:R_indirect}. Noticeably, in Figure \ref{fig:indirect_1_nonconformity_distribution}, we see that $C\alpha_\tau$ is less conservative than $C$ from \cite{lindemann2023conformal}. \emph{Variant II.} In Figure \ref{fig:indirect_2_nonconformity_distribution}, we show the histogram of $R^{(i)}$ in equation \eqref{eq:R_hybrid} over calibration data along with $\tilde{C}$ from \eqref{eq:C_tilde} and $C$ from \eqref{eq:vanilla_quantile}. As expected, the indirect algorithms are more conservative than the direct algorithm and all achieve an empirical coverage for $\rho^\phi(X^{(i)}, \tau_0) \ge \rho^*$ greater than the desired coverage of $1 - \delta = 0.8$. This is demonstrated in Figure \ref{fig:case_study_indirect_coverage} where we plot the coverage over the $50$ experiments. For one experiment, we show the true robust semantics $\rho^\phi(X, \tau_0)$ for the 200 ground truth test data  and the predicted worst-case robust semantics $\rho^*$ for Variants I and II in Figure \ref{fig:case_study_indirect_robustnesses}. Finally, in comparison with the non-robust indirect method from \cite{lindemann2023conformal}, we see that $\rho^*$ attained by Variant I (green) and Variant II (purple) are higher than that from \cite{lindemann2023conformal} (orange) further indicating a reduction in conservatism.

\section{Conclusion}
\label{sec:conclusion}

We presented two predictive runtime verification algorithms that can predict system failures even when the test time system is different from the design time system. Specifically, our algorithms are robust against distribution shifts that are measured in terms of the $f$-divergence of their system trajectories. We first use trajectory predictors to predict the future motion of the system, and we robustly quantify prediction uncertainty with respect to signal temporal logic (STL) system specifications using robust conformal prediction and calibration data from the design time system. Our first algorithm (called direct algorithm) provides tight  verification guarantees, while our second algorithm (called indirect algorithm), which we present in two variants, provides more interpretable runtime information. Finally, we  quantify
the relationship between calibration data, desired confidence,
and permissible distribution shift. Both algorithms make no assumption on the trajectory predictor and only need to know an upper bound on the distribution shift. To the best of our knowledge, these are the first results in this setting. We provide an exhaustive validation of our algorithms on a Franka robot within NVIDIA Isaac Sim. 

\section{Acknowledgements}
This work was partially supported by the National Science Foundation through the following grants: CAREER award (SHF-2048094), CNS-1932620, CNS-2039087, FMitF-1837131, CCF-SHF-1932620, funding by Toyota R\&D and Siemens Corporate Research through the USC Center for Autonomy and AI, an Amazon Faculty Research Award, and the Airbus Institute for Engineering Research. This work does not reflect the views of any organization listed.


\bibliographystyle{IEEEtran}
\bibliography{main}

\begin{thebibliography}{10}
\providecommand{\url}[1]{#1}
\csname url@samestyle\endcsname
\providecommand{\newblock}{\relax}
\providecommand{\bibinfo}[2]{#2}
\providecommand{\BIBentrySTDinterwordspacing}{\spaceskip=0pt\relax}
\providecommand{\BIBentryALTinterwordstretchfactor}{4}
\providecommand{\BIBentryALTinterwordspacing}{\spaceskip=\fontdimen2\font plus
\BIBentryALTinterwordstretchfactor\fontdimen3\font minus \fontdimen4\font\relax}
\providecommand{\BIBforeignlanguage}[2]{{%
\expandafter\ifx\csname l@#1\endcsname\relax
\typeout{** WARNING: IEEEtran.bst: No hyphenation pattern has been}%
\typeout{** loaded for the language `#1'. Using the pattern for}%
\typeout{** the default language instead.}%
\else
\language=\csname l@#1\endcsname
\fi
#2}}
\providecommand{\BIBdecl}{\relax}
\BIBdecl

\bibitem{lindemann2023conformal}
L.~Lindemann, X.~Qin, J.~V. Deshmukh, and G.~J. Pappas, ``Conformal prediction for stl runtime verification,'' in \emph{Proc. of ICCPS}, 2023, pp. 142--153.

\bibitem{hekmatnejad2019encoding}
M.~Hekmatnejad, S.~Yaghoubi, A.~Dokhanchi, H.~B. Amor, A.~Shrivastava, L.~Karam, and G.~Fainekos, ``Encoding and monitoring responsibility sensitive safety rules for automated vehicles in signal temporal logic,'' in \emph{Proc. of MEMOCODE}, 2019, pp. 1--11.

\bibitem{cairoli2023conformal}
F.~Cairoli, N.~Paoletti, and L.~Bortolussi, ``Conformal quantitative predictive monitoring of stl requirements for stochastic processes,'' in \emph{Proc. of HSCC}, 2023, pp. 1--11.

\bibitem{cauchois2020robust}
M.~Cauchois, S.~Gupta, A.~Ali, and J.~C. Duchi, ``Robust validation: Confident predictions even when distributions shift,'' \emph{arXiv preprint arXiv:2008.04267}, 2020.

\bibitem{asarin1998achilles}
E.~Asarin and O.~Maler, ``Achilles and the tortoise climbing up the arithmetical hierarchy,'' \emph{Journal of Computer and System Sciences}, vol.~57, no.~3, pp. 389--398, 1998.

\bibitem{agha2018survey}
G.~Agha and K.~Palmskog, ``A survey of statistical model checking,'' \emph{ACM Transactions on Modeling and Computer Simulation (TOMACS)}, vol.~28, no.~1, pp. 1--39, 2018.

\bibitem{legay2019statistical}
A.~Legay, A.~Lukina, L.~M. Traonouez, J.~Yang, S.~A. Smolka, and R.~Grosu, ``Statistical model checking,'' in \emph{Computing and software science: state of the art and perspectives}.\hskip 1em plus 0.5em minus 0.4em\relax Springer, 2019, pp. 478--504.

\bibitem{zarei2020statistical}
M.~Zarei, Y.~Wang, and M.~Pajic, ``Statistical verification of learning-based cyber-physical systems,'' in \emph{Proc. Int. Conf. on Hybrid Syst.: Comp. and Control}, 2020, pp. 1--7.

\bibitem{sen2004statistical}
K.~Sen, M.~Viswanathan, and G.~Agha, ``Statistical model checking of black-box probabilistic systems,'' in \emph{Proc. of CAV}.\hskip 1em plus 0.5em minus 0.4em\relax Springer, 2004, pp. 202--215.

\bibitem{bartocci2013robustness}
E.~Bartocci, L.~Bortolussi, L.~Nenzi, and G.~Sanguinetti, ``On the robustness of temporal properties for stochastic models,'' in \emph{Proc. Int. Workshop Hybrid Syst. Biology}, Taormina, Italy, Sept. 2013, pp. 3--19.

\bibitem{bartocci2015system}
------, ``System design of stochastic models using robustness of temporal properties,'' \emph{Theoret. Comp. Science}, vol. 587, pp. 3--25, 2015.

\bibitem{qin2022statistical}
X.~Qin, Y.~Xian, A.~Zutshi, C.~Fan, and J.~V. Deshmukh, ``Statistical verification of cyber-physical systems using surrogate models and conformal inference,'' in \emph{Proc. of ICCPS}, May 2022, pp. 116--126.

\bibitem{salamati2021data}
A.~Salamati, S.~Soudjani, and M.~Zamani, ``Data-driven verification of stochastic linear systems with signal temporal logic constraints,'' \emph{Automatica}, vol. 131, p. 109781, 2021.

\bibitem{salamati2020data}
------, ``Data-driven verification under signal temporal logic constraints,'' \emph{IFAC-PapersOnLine}, vol.~53, no.~2, pp. 69--74, 2020.

\bibitem{pedrielli2023part}
G.~Pedrielli, T.~Khandait, Y.~Cao, Q.~Thibeault, H.~Huang, M.~Castillo-Effen, and G.~Fainekos, ``Part-x: A family of stochastic algorithms for search-based test generation with probabilistic guarantees,'' \emph{IEEE Transactions on Automation Science and Engineering}, 2023.

\bibitem{dutta2023distributionally}
S.~Dutta, M.~Caprio, V.~Lin, M.~Cleaveland, K.~J. Jang, I.~Ruchkin, O.~Sokolsky, and I.~Lee, ``Distributionally robust statistical verification with imprecise neural networks,'' \emph{arXiv preprint arXiv:2308.14815}, 2023.

\bibitem{wu2023toward}
H.~Wu, T.~Tagomori, A.~Robey, F.~Yang, N.~Matni, G.~Pappas, H.~Hassani, C.~Pasareanu, and C.~Barrett, ``Toward certified robustness against real-world distribution shifts,'' in \emph{2023 IEEE Conference on Secure and Trustworthy Machine Learning (SaTML)}.\hskip 1em plus 0.5em minus 0.4em\relax IEEE, 2023, pp. 537--553.

\bibitem{wong2020learning}
E.~Wong and J.~Z. Kolter, ``Learning perturbation sets for robust machine learning,'' \emph{arXiv preprint arXiv:2007.08450}, 2020.

\bibitem{bartocci2018introduction}
E.~Bartocci, Y.~Falcone, A.~Francalanza, and G.~Reger, ``Introduction to runtime verification,'' \emph{Lectures on Runtime Verification: Introductory and Advanced Topics}, pp. 1--33, 2018.

\bibitem{bauer2011runtime}
A.~Bauer, M.~Leucker, and C.~Schallhart, ``Runtime verification for ltl and tltl,'' \emph{ACM Transactions on Software Engineering and Methodology (TOSEM)}, vol.~20, no.~4, pp. 1--64, 2011.

\bibitem{lukina2021into}
A.~Lukina, C.~Schilling, and T.~A. Henzinger, ``Into the unknown: Active monitoring of neural networks,'' in \emph{International Conference on Runtime Verification}.\hskip 1em plus 0.5em minus 0.4em\relax Springer, 2021, pp. 42--61.

\bibitem{jaeger2020statistical}
M.~Jaeger, K.~G. Larsen, and A.~Tibo, ``From statistical model checking to run-time monitoring using a bayesian network approach,'' in \emph{International Conference on Runtime Verification}.\hskip 1em plus 0.5em minus 0.4em\relax Springer, 2020, pp. 517--535.

\bibitem{ma2021predictive}
M.~Ma, J.~Stankovic, E.~Bartocci, and L.~Feng, ``Predictive monitoring with logic-calibrated uncertainty for cyber-physical systems,'' \emph{ACM Transactions on Embedded Computing Systems (TECS)}, vol.~20, no.~5s, pp. 1--25, 2021.

\bibitem{deshmukh2017robust}
J.~V. Deshmukh, A.~Donz{\'e}, S.~Ghosh, X.~Jin, G.~Juniwal, and S.~A. Seshia, ``Robust online monitoring of signal temporal logic,'' \emph{Formal Methods in System Design}, vol.~51, no.~1, pp. 5--30, 2017.

\bibitem{dokhanchi2014line}
A.~Dokhanchi, B.~Hoxha, and G.~Fainekos, ``On-line monitoring for temporal logic robustness,'' in \emph{RV}.\hskip 1em plus 0.5em minus 0.4em\relax Springer, 2014, pp. 231--246.

\bibitem{gressenbuch2021predictive}
L.~Gressenbuch and M.~Althoff, ``Predictive monitoring of traffic rules,'' in \emph{Proc. of ITSC}.\hskip 1em plus 0.5em minus 0.4em\relax IEEE, 2021, pp. 915--922.

\bibitem{sistla2008monitoring}
A.~P. Sistla and A.~R. Srinivas, ``Monitoring temporal properties of stochastic systems,'' in \emph{International Workshop on Verification, Model Checking, and Abstract Interpretation}.\hskip 1em plus 0.5em minus 0.4em\relax Springer, 2008, pp. 294--308.

\bibitem{sistla2011runtime}
A.~P. Sistla, M.~{\v{Z}}efran, and Y.~Feng, ``Runtime monitoring of stochastic cyber-physical systems with hybrid state,'' in \emph{International Conference on Runtime Verification}.\hskip 1em plus 0.5em minus 0.4em\relax Springer, 2011, pp. 276--293.

\bibitem{bortolussi2019neural}
L.~Bortolussi, F.~Cairoli, N.~Paoletti, S.~A. Smolka, and S.~D. Stoller, ``Neural predictive monitoring,'' in \emph{International Conference on Runtime Verification}.\hskip 1em plus 0.5em minus 0.4em\relax Springer, 2019, pp. 129--147.

\bibitem{qin2020clairvoyant}
X.~Qin and J.~V. Deshmukh, ``Clairvoyant monitoring for signal temporal logic,'' in \emph{International Conference on Formal Modeling and Analysis of Timed Systems}.\hskip 1em plus 0.5em minus 0.4em\relax Springer, 2020, pp. 178--195.

\bibitem{cairoli2023learning}
F.~Cairoli, L.~Bortolussi, and N.~Paoletti, ``Learning-based approaches to predictive monitoring with conformal statistical guarantees,'' in \emph{International Conference on Runtime Verification}.\hskip 1em plus 0.5em minus 0.4em\relax Springer, 2023, pp. 461--487.

\bibitem{yu2024modeljournal}
X.~Yu, W.~Dong, S.~Li, and X.~Yin, ``Model predictive monitoring of dynamical systems for stl specifications,'' \emph{Automatica}, 2024.

\bibitem{yu2022online}
X.~Yu, W.~Dong, X.~Yin, and S.~Li, ``Online monitoring of dynamic systems for signal temporal logic specifications with model information,'' in \emph{2022 IEEE 61st Conference on Decision and Control (CDC)}.\hskip 1em plus 0.5em minus 0.4em\relax IEEE, 2022, pp. 1553--1559.

\bibitem{wang2023sleep}
C.~Wang, X.~Yu, J.~Zhao, L.~Lindemann, and X.~Yin, ``Sleep when everything looks fine: Self-triggered monitoring for signal temporal logic tasks,'' \emph{arXiv preprint arXiv:2311.15531}, 2023.

\bibitem{vovk2005algorithmic}
V.~Vovk, A.~Gammerman, and G.~Shafer, \emph{Algorithmic learning in a random world}.\hskip 1em plus 0.5em minus 0.4em\relax Springer Science \& Business Media, 2005.

\bibitem{romano2019conformalized}
Y.~Romano, E.~Patterson, and E.~Candes, ``Conformalized quantile regression,'' \emph{Advances in neural information processing systems}, vol.~32, 2019.

\bibitem{mao2023safe}
Z.~Mao, C.~Sobolewski, and I.~Ruchkin, ``How safe am i given what i see? calibrated prediction of safety chances for image-controlled autonomy,'' \emph{arXiv preprint arXiv:2308.12252}, 2023.

\bibitem{maler2004monitoring}
O.~Maler and D.~Nickovic, ``Monitoring temporal properties of continuous signals,'' in \emph{Proc. Int. Conf. FORMATS FTRTFT}, Grenoble, France, September 2004, pp. 152--166.

\bibitem{donze2}
A.~Donz{\'e} and O.~Maler, ``Robust satisfaction of temporal logic over real-valued signals,'' in \emph{Proc. Int. Conf. FORMATS}, Klosterneuburg, Austria, September 2010, pp. 92--106.

\bibitem{fainekos2009robustness}
G.~E. Fainekos and G.~J. Pappas, ``Robustness of temporal logic specifications for continuous-time signals,'' \emph{Theoret. Comp. Science}, vol. 410, no.~42, pp. 4262--4291, 2009.

\bibitem{sadraddini2015robust}
S.~Sadraddini and C.~Belta, ``Robust temporal logic model predictive control,'' in \emph{Proceedings of the 53rd Annual Allerton Conference on Communication, Control, and Computing}, Monticello, IL, September 2015, pp. 772--779.

\bibitem{heidlauf2018verification}
P.~Heidlauf, A.~Collins, M.~Bolender, and S.~Bak, ``Verification challenges in f-16 ground collision avoidance and other automated maneuvers.'' in \emph{ARCH@ ADHS}, 2018, pp. 208--217.

\bibitem{shafer2008tutorial}
G.~Shafer and V.~Vovk, ``A tutorial on conformal prediction.'' \emph{Journal of Machine Learning Research}, vol.~9, no.~3, 2008.

\bibitem{angelopoulos2021gentle}
A.~N. Angelopoulos and S.~Bates, ``A gentle introduction to conformal prediction and distribution-free uncertainty quantification,'' \emph{arXiv preprint arXiv:2107.07511}, 2021.

\bibitem{lei2018distribution}
J.~Lei, M.~G’Sell, A.~Rinaldo, R.~J. Tibshirani, and L.~Wasserman, ``Distribution-free predictive inference for regression,'' \emph{Journal of the American Statistical Association}, vol. 113, no. 523, pp. 1094--1111, 2018.

\bibitem{hochreiter1997long}
S.~Hochreiter and J.~Schmidhuber, ``Long short-term memory,'' \emph{Neural computation}, vol.~9, no.~8, pp. 1735--1780, 1997.

\bibitem{fjellstrom2022long}
C.~Fjellstr$\ddot{o}$m, ``Long short-term memory neural network for financial time series,'' in \emph{Proceedings of 2022 International Conference on Big Data}.\hskip 1em plus 0.5em minus 0.4em\relax IEEE, 2022, pp. 3496--3504.

\bibitem{cortes1995support}
C.~Cortes and V.~Vapnik, ``Support-vector networks,'' \emph{Machine learning}, vol.~20, pp. 273--297, 1995.

\bibitem{ristanoski2013time}
G.~Ristanoski, W.~Liu, and J.~Bailey, ``A time-dependent enhanced support vector machine for time series regression,'' in \emph{Proc. SIGKDD int. conf. on Knowledge discovery and data mining}, 2013, pp. 946--954.

\bibitem{box2015time}
G.~E. Box, G.~M. Jenkins, G.~C. Reinsel, and G.~M. Ljung, \emph{Time series analysis: forecasting and control}.\hskip 1em plus 0.5em minus 0.4em\relax John Wiley \& Sons, 2015.

\bibitem{mehrmolaei2016time}
S.~Mehrmolaei and M.~R. Keyvanpour, ``Time series forecasting using improved arima,'' in \emph{2016 Artificial Intelligence and Robotics (IRANOPEN)}.\hskip 1em plus 0.5em minus 0.4em\relax IEEE, 2016, pp. 92--97.

\bibitem{cleaveland2023lcp}
M.~Cleaveland, I.~Lee, G.~J. Pappas, and L.~Lindemann, ``Conformal prediction regions for time series using linear complementarity programming,'' \emph{arXiv preprint arXiv:2304.01075}, 2023.

\bibitem{makoviychuk2021isaac}
V.~Makoviychuk, L.~Wawrzyniak, Y.~Guo, M.~Lu, K.~Storey, M.~Macklin, D.~Hoeller, N.~Rudin, A.~Allshire, A.~Handa \emph{et~al.}, ``Isaac gym: High performance gpu-based physics simulation for robot learning,'' \emph{arXiv preprint arXiv:2108.10470}, 2021.

\bibitem{zhou1996robust}
J.~Doyle, ``Robust and optimal control,'' \emph{Control Engineering Practice}, vol.~4, no.~8, pp. 1189--1190, 1996.

\bibitem{lindemann2019robust}
L.~Lindemann and D.~V. Dimarogonas, ``Robust control for signal temporal logic specifications using discrete average space robustness,'' \emph{Automatica}, vol. 101, pp. 377--387, 2019.

\end{thebibliography}

\addtolength{\textheight}{-2cm}   

\newpage
\appendix
\begin{center}
\section*{Appendix}
\end{center}

\section{Semantics of Signal Temporal Logic}
\label{app:STL}
For a trajectory $x:=(x_0,x_1,\hdots)$, the semantics of an STL formula $\phi$ that is enabled at time $\tau_0$, denoted by $(x,\tau_0)\models \phi$, can be recursively computed based on the structure of $\phi$ using the following rules:
	\begin{align*}
	(x,\tau)\models \text{True} & \hspace{0.5cm} \text{iff} \hspace{0.5cm} \text{True},\\
	(x,\tau)\models \pi & \hspace{0.5cm} \text{iff} \hspace{0.5cm} h(x_\tau)\ge 0,\\
	(x,\tau)\models \neg\phi & \hspace{0.5cm} \text{iff} \hspace{0.5cm} (x,\tau)\not\models \phi,\\
	(x,\tau)\models \phi' \wedge \phi'' & \hspace{0.5cm} \text{iff} \hspace{0.5cm} (x,\tau)\models\phi' \text{ and } (x,\tau)\models\phi'',\\
	(x,\tau)\models \phi' U_I \phi'' & \hspace{0.5cm} \text{iff} \hspace{0.5cm} \exists \tau''\in (\tau\oplus I)\cap \mathbb{N} \text{ s.t. } (x,\tau'')\models\phi''\\
	&\hspace{1.2cm} \text{ and } \forall \tau'\in(\tau,\tau'')\cap \mathbb{N}, (x,\tau')\models\phi'.
	\end{align*}

The robust semantics $\rho^{\phi}(x,\tau_0)$ provide more information than the semantics $(x,\tau_0)\models \phi$, and indicate how robustly a specification is satisfied. We can again recursively calculate $\rho^{\phi}(x,\tau_0)$ based on the structure of $\phi$ using the following rules:
\begin{align*}
	\rho^\text{True}(x,\tau)& := \infty,\\
	\rho^{\pi}(x,\tau)& := h(x_\tau) \\
	\rho^{\neg\phi}(x,\tau) &:= 	-\rho^{\phi}(x,\tau),\\
	\rho^{\phi' \wedge \phi''}(x,\tau) &:= 	\min(\rho^{\phi'}(x,\tau),\rho^{\phi''}(x,\tau)),\\
	\rho^{\phi' U_I \phi''}(x,\tau) &:= \underset{\tau''\in (\tau\oplus I)\cap \mathbb{N}}{\text{sup}}  \Big(\min\big(\rho^{\phi''}(x,\tau''),\underset{\tau'\in (\tau,\tau'')\cap \mathbb{N}}{\text{inf}}\rho^{\phi'}(x,\tau') \big)\Big).
	\end{align*}
	
	 The formula length $L^{\phi}$ of a bounded STL formula $\phi$ can be recursively calculated based on the structure of $\phi$ using the following rules:
 \begin{align*}
     L^\text{True}&=L^\pi:=0\\
     L^{\neg\phi}&:=L^\phi\\
     L^{\phi'\wedge\phi''}&:=\max(L^{\phi'},L^{\phi''})\\
     L^{\phi' U_I \phi''}&:=\max \{I\cap \mathbb{N}\}+\max(L^{\phi'},L^{\phi''}).
 \end{align*}

 The probabilistic robust semantics $\bar{\rho}^{\phi}(X,\tau_0)$, which are defined over the random trajectory $X\sim \mathcal{D}$ and the predicted trajectory $\hat{X} := (X_\text{obs}, \hat{X}_{t + 1 | t} , \hdots, \hat{X}_{t + H | t})$, are recursively defined based on the structure of $\phi$ using the following rules:
\begin{align*}
    \bar{\rho}^\text{True}(\hat{X},\tau)& := \infty,\\
    \bar{\rho}^{\pi}(\hat{X},\tau)& := 
     \begin{cases}
     h(X_\tau) &\text{ if } \tau\le t\\
     \rho_{\pi,\tau}^* &\text{ otherwise }
     \end{cases}\\
	\bar{\rho}^{\phi' \wedge \phi''}(\hat{X},\tau) &:= 	\min(\bar{\rho}^{\phi'}(\hat{X},\tau),\bar{\rho}^{\phi''}(\hat{X},\tau)),\\
	\bar{\rho}^{\phi' U_I \phi''}(\hat{X},\tau) &:= \underset{\tau''\in (\tau\oplus I)\cap \mathbb{N}}{\text{sup}}  \Big(\min\big(\bar{\rho}^{\phi''}(\hat{X},\tau''),\underset{\tau'\in (\tau,\tau'')\cap \mathbb{N}}{\text{inf}}\bar{\rho}^{\phi'}(\hat{X},\tau') \big)\Big)
	\end{align*}
 where the constant $\rho_{\pi,\tau}^*$ defines probabilistic prediction regions that can be computed as explained in detail in Section \ref{sec:f_div}.

\section{Proofs for Technical Theorems and Lemmas}
\label{app:proof}
\subsection{Proof of Theorem \ref{theorem:1}}
\begin{proof}
 By the data processing inequality in Lemma \ref{lemma:2} and since $D_f(\mathcal{D},\mathcal{D}_0)\le \epsilon$ holds by Assumption \ref{ass2}, we know that $D(\mathcal{R},\mathcal{R}_0)\le \epsilon$. We can thus apply Lemma \ref{lemma:1} and construct $\tilde{C}$ according to equation \eqref{eq:C_tilde} with $R^{(i)}$ as in \eqref{eq:R_direct}. We then know that $Prob(\rho^\phi(\hat{X}, \tau_0) - \rho^\phi(X, \tau_0) \le \tilde{C}) \ge 1 - \delta$. Therefore, it follows that $Prob(\rho^\phi(X, \tau_0) \ge \rho^\phi(\hat{X}, \tau_0) - \tilde{C}) \ge 1 - \delta$.
\end{proof}

\subsection{Proof of Theorem \ref{theorem:2}}

\begin{proof}
By assumption we know that equation \eqref{eq:guarantee_prediate} holds, i.e., that $\text{Prob}(\rho^{\pi}(X, \tau) \ge \rho_{\pi,\tau}^*, \forall (\pi, \tau) \in \mathcal{P}) \ge 1 - \delta$. Since we define $\bar{\rho}^{\pi}(\hat{X},\tau) :=h(X_\tau)$ if $\tau\le t$ and $\bar{\rho}^{\pi}(\hat{X},\tau) :=\rho_{\pi,\tau}^*$ otherwise, we know that $\text{Prob}(\rho^{\pi}(X, \tau) \ge \bar{\rho}^{\pi}(\hat{X},\tau), \forall (\pi, \tau) \in \mathcal{P}) \ge 1 - \delta$. Since $\phi$ is in positive normal form, we further know that, for any two signals $y,y':\mathbb{N}\to \mathbb{R}^n$, it holds that ${\rho}^{\pi}(y, \tau) \ge {\rho}^{\pi}(y', \tau)$ for all $(\pi, \tau) \in \mathcal{P}$ implies ${\rho}^{\phi}(y, \tau_0) \ge {\rho}^{\phi}(y', \tau_0)$ \cite[Corollary 1]{lindemann2019robust}. Consequently, we conclude that $\text{Prob}(\rho^\phi(X, \tau_0) \ge \bar{\rho}^\phi(\hat{X}, \tau_0)) \ge 1 - \delta$ since the Boolean and temporal operators for $\bar{\rho}^\phi$ follow the same semantics as for ${\rho}^\phi$.
\end{proof}

\subsection{Proof of Lemma \ref{lemma:variant1}}

\begin{proof}
     By the data processing inequality in Lemma \ref{lemma:2} and since $D_f(\mathcal{D},\mathcal{D}_0)\le \epsilon$ holds by Assumption \ref{ass2}, we know that $D(\mathcal{R},\mathcal{R}_0)\le \epsilon$. We can thus apply Lemma \ref{lemma:1} and construct $\tilde{C}$ according to equation \eqref{eq:C_tilde} with $R^{(i)}$ as in \eqref{eq:R_indirect}. We then know that $\text{Prob}(\max_{\tau \in \{t+1, \hdots, t+H\}} \frac{\|X_\tau - \hat{X}_{\tau|t}\|}{\alpha_\tau} \le \tilde{C}) \ge 1 - \delta$, which implies that $\text{Prob}(\frac{\|X_\tau - \hat{X}_{\tau|t}\|}{\alpha_\tau} \le \tilde{C}, \forall \tau \in \{t + 1, \hdots, t + H\}) \ge 1 - \delta$. Since  $\alpha_\tau > 0$, this is equivalent to  $\text{Prob}(\|X_\tau - \hat{X}_{\tau|t}\|\le \tilde{C}\alpha_\tau,  \forall \tau \in \{t+1, \hdots, t+H\}) \ge 1 - \delta$. Finally, by the definition of $\rho^*_{\pi, \tau}$ in equation \eqref{eq:worst_case_robustness}, which considers the worst case value of $h(\zeta)$ over $\zeta\in \mathcal{B}_{\tau}$, it follows that  $\text{Prob}(\rho^{\pi}(X, \tau) \ge \rho_{\pi,\tau}^*, \forall (\pi, \tau) \in \mathcal{P}) \ge 1 - \delta$ holds.
\end{proof}

\subsection{Proof of Lemma \ref{lemma:variant2}}

\begin{proof}
 By the data processing inequality in Lemma \ref{lemma:2} and since $D_f(\mathcal{D},\mathcal{D}_0)\le \epsilon$ holds by Assumption \ref{ass2}, we know that $D(\mathcal{R},\mathcal{R}_0)\le \epsilon$. We can thus apply Lemma \ref{lemma:1} and construct $\tilde{C}$ according to equation \eqref{eq:C_tilde} with $R^{(i)}$ as in \eqref{eq:R_hybrid}. We then know that $\text{Prob}(\max_{(\pi, \tau) \in \mathcal{P}}\frac{\rho^{\pi}(\hat{X}, \tau) - \rho^{\pi}(X, \tau)}{\alpha_{\pi, \tau}} \le \tilde{C}) \ge 1 - \delta$, which implies that  $\text{Prob}(\frac{\rho^{\pi}(\hat{X}, \tau) - \rho^{\pi}(X, \tau)}{\alpha_{\pi, \tau}} \le \tilde{C}, \forall (\pi, \tau) \in \mathcal{P}) \ge 1 - \delta$. Since $\alpha_{\pi, \tau} > 0$, this is equivalent to $\text{Prob}(\rho^{\pi}(\hat{X}, \tau) - \rho^{\pi}(X, \tau) \le \tilde{C}\alpha_{\pi, \tau}, \forall (\pi, \tau) \in \mathcal{P}) \ge 1 - \delta$. From here, it follows that $\text{Prob}(\rho^{\pi}(X, \tau) \ge \rho_{\pi,\tau}^*, \forall (\pi, \tau) \in \mathcal{P}) \ge 1 - \delta$  with $\rho_{\pi,\tau}^*:= \rho^{\pi}(\hat{X}, \tau)-\tilde{C}\alpha_{\pi, \tau}$.
\end{proof}
\end{document}